\documentclass[12pt]{article}
\usepackage{a4wide,amssymb,epsfig,color}
\usepackage{epsfig,graphicx,float,authordate1-4}
\usepackage{amsmath}
\usepackage[tight]{subfigure}

\newcommand{\Pc}{\mbox{$P^\mathrm{C}$}}

\newcommand{\D}{\mbox{$\,\mathrm{d}$}}
\newcommand{\Dx}{\mbox{$\mathrm{d}$}}

\newcommand{\mm}{\mbox{$\mathrm{mm}$}}

\newcommand{\Pco}{\mbox{$P_o^\mathrm{C}$}}

\DeclareMathOperator{\sech}{sech}

\begin{document}

\title{Local--global mode interaction\\
  in stringer-stiffened plates\footnote{Article accepted for
    \emph{Thin-Walled Structures}, 19 September 2014.}}

\author{M. Ahmer Wadee and Maryam Farsi\\~\\
  Department of Civil \& Environmental Engineering,\\
  Imperial College London,\\
  London SW7 2AZ, UK}

\date{}

\maketitle

\begin{abstract}

  A recently developed nonlinear analytical model for axially loaded
  thin-walled stringer-stiffened plates based on variational
  principles is extended to include local buckling of the main
  plate. Interaction between the weakly stable global buckling mode
  and the strongly stable local buckling mode is highlighted. Highly
  unstable post-buckling behaviour and a progressively changing
  wavelength in the local buckling mode profile is observed under
  increasing compressive deformation. The analytical model is compared
  against both physical experiments from the literature and finite
  element analysis conducted in the commercial code \textsc{Abaqus};
  excellent agreement is found both in terms of the mechanical
  response and the predicted deflections.

\end{abstract}

\subsubsection*{Key words}
  Mode interaction; Stiffened plates; Cellular buckling; Snaking;
  Nonlinear mechanics.

\section{Introduction}
\label{sec:intro}

Thin-walled stringer-stiffened plates under axial compression are well
known to be vulnerable to buckling where local and global modes
interact nonlinearly \cite{Koiter1976,Fok1976,IUTAM76,TH84}. However,
since stiffened plates are highly mass-efficient structural
components, their application is ubiquitous in long-span bridge decks
\cite{Ronalds89}, ships and offshore structures \cite{Murray1973}, and
aerospace structures \cite{Butler2000,loughlan2004thin}. Hence,
understanding the behaviour of these components represents a
structural problem of enormous practical significance
\cite{Grondin1999,Sheikh2002,Ghavami2006}. Other significant
structural components such as sandwich struts \cite{HW1998}, built-up
columns \cite{TH73}, corrugated plates \cite{Pignataro2000} and other
thin-walled components
\cite{Hancock1981,Schafer2002,Becque2009a,WB2014} are also well-known
to suffer from the instabilities arising from the interaction of
global and local buckling modes.

In the authors' recent work \cite{WF1_2014}, the aforementioned
problem was studied using an analytical approach by considering that
interactive buckling was wholly confined to the stringer (or stiffener)
only. So-called ``cellular buckling'' \cite{Hunt2000,WG2012,WB2014,BW2014} or
``snaking'' \cite{WC1999,BurkeKnobloch2007,CK2009} was captured, where
snap-backs in the response, showing sequential destabilization and
restabilization and a progressive spreading of the initial localized
buckling mode, were revealed. The results showed reasonably good
comparisons with a finite element (FE) model formulated in the
commercial code \textsc{Abaqus} \shortcite{ABAQUS}.  The current work
extends the previous model such that the interaction between global
Euler buckling and the local buckling of the main plate, as well as
the stiffener, are accounted. A system of nonlinear ordinary
differential equations subject to integral constraints is derived
using variational principles and is subsequently solved using the
numerical continuation package \textsc{Auto-07p} \cite{auto}. The
relative rigidity of the main plate--stiffener joint is adjusted by
means of a rotational spring, increasing the stiffness of which
results in the erosion of the snap-backs that signify cellular
buckling.  However, the changing local buckling wavelength is still
observed, although the effect is not quite so marked as compared with
the case where the joint is assumed to be pinned \cite{WF1_2014}. A
finite element model is also developed using the commercial code
\textsc{Abaqus} for validation purposes. Moreover, given that local
buckling of the main plate is included alongside the buckling of the
stiffener in the current model, which is often observed in
experiments, the present results are also compared with a couple of
physical test results from the literature \cite{Fok1976}. The
comparisons turn out to be excellent both in terms of the mechanical
response and the physical post-buckling profiles.

\section{Analytical Model}

\begin{figure}[htb]
  \centerline{\includegraphics[width=130mm]{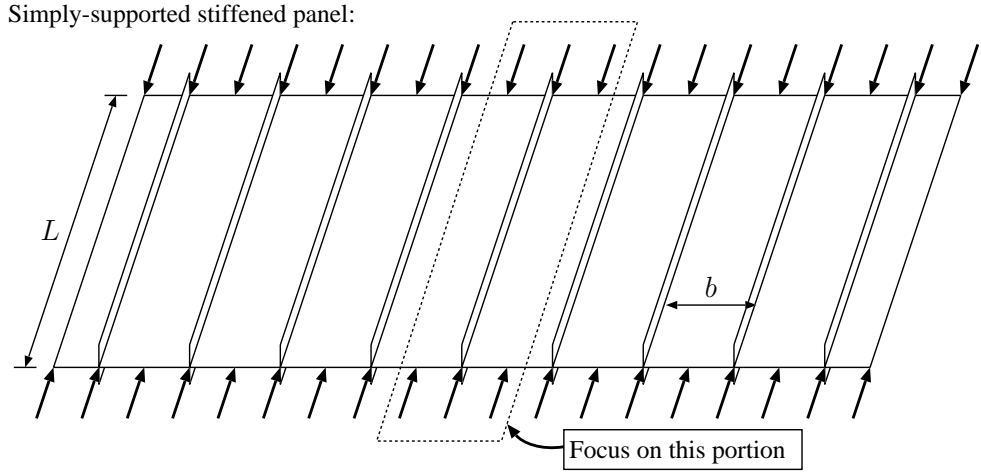}}
  \caption{An axially compressed simply-supported stiffened panel of
    length $L$ and evenly spaced stiffeners separated by a distance
    $b$.}
\label{fig:panel}
\end{figure}

Consider a thin-walled simply-supported plated panel that has
uniformly spaced stiffeners above and below the main plate, as shown
in Figure \ref{fig:panel},
with panel length $L$ and the spacing between the stiffeners being
$b$. It is made from a linear elastic, homogeneous and isotropic
material with Young's modulus $E$, Poisson's ratio $\nu$ and shear
modulus $G=E/[2(1+\nu)]$. If the panel is much wider than long,
\emph{i.e.}\ $L \ll n_s b$, where $n_s$ is the number of stiffeners in
the panel, the critical buckling behaviour of the panel would be
strut-like with a half-sine wave eigenmode along the length. Moreover,
this would allow a portion of the panel that is representative of its
entirety to be isolated as a strut as depicted in Figure
\ref{fig:panel}, since the transverse bending curvature of the panel
during initial post-buckling would be relatively small.

\begin{figure}[hbt]
\centering
\subfigure[]{\includegraphics[scale=1.1]{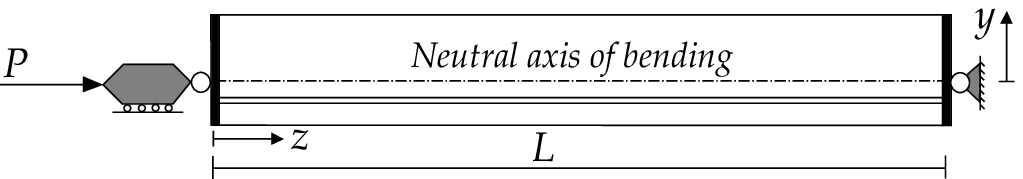}}\\
\subfigure[]{\includegraphics[scale=1.0]{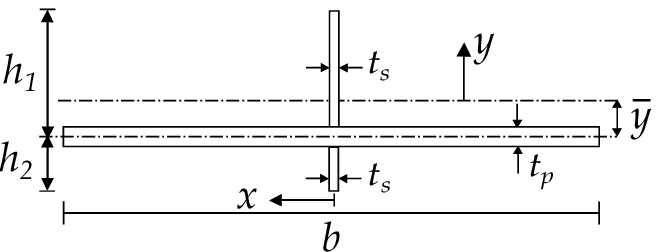}}
\subfigure[]{\includegraphics[scale=1.0]{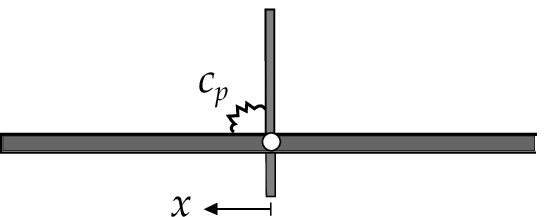}}
\subfigure[]{\includegraphics[scale=0.75]{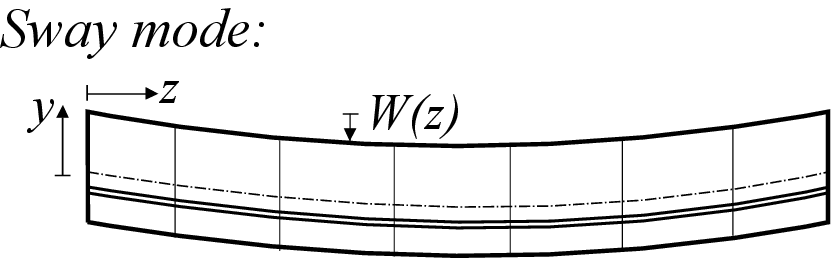}\qquad
  \includegraphics[scale=0.75]{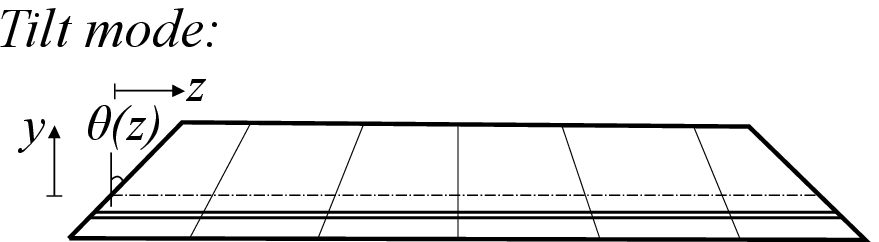}}
\caption{(a) Elevation of the representative portion of the stiffened
  plate modelled as strut of length $L$ that is compressed axially by
  a force $P$. (b) Strut cross-section geometry. (c) Modelling the
  joint rigidity of the main plate--stiffener connection with a
  rotational spring of stiffness $c_p$. (d) Sway and tilt components
  of the global buckling mode.}
\label{fig:cross_section}
\end{figure}

Therefore, the current article presents an analytical model of a
representative portion of an axially-compressed stiffened panel, which
simplifies to a simply supported strut with geometric properties
defined in Figure \ref{fig:cross_section}.
The strut has length $L$ and comprises a main plate (or skin) of width
$b$ and thickness $t_p$ with two attached longitudinal stiffeners of
heights $h_1$ and $h_2$ with thickness $t_s$, as shown in Figure
\ref{fig:cross_section}(b). The axial load $P$ is applied at the
centroid of the whole cross-section denoted as the distance $\bar{y}$
from the centre line of the plate. The rigidity of the connection
between the main plate and stiffeners is modelled with a rotational
spring of stiffness $c_p$, as shown in Figure
\ref{fig:cross_section}(c). If $c_p = 0$, a pinned joint is modelled,
but if $c_p$ is large, the joint is considered to be completely fixed
or rigid. Note that the rotational spring with stiffness $c_p$ only
stores strain energy by local bending of the stiffener or the main
plate at the joint coordinates ($x=0,y=-\bar{y}$) and not by rigid
body rotation of the entire joint in a twisting action.

\subsection{Modal descriptions}

To model interactive buckling analytically, it has been demonstrated
that shear strains need to be included \cite{asmejam98,WYT2010} and for
thin-walled metallic elements Timoshenko beam theory has been shown to
be sufficiently accurate \cite{WG2012,WB2014}. To model the global
buckling mode, two degrees of freedom, known as ``sway'' and ``tilt''
in the literature \cite{HDSM88}, are used. The sway mode is
represented by the displacement $W$ of the plane sections that are
under global flexure and the tilt mode is represented by the
corresponding angle of inclination $\theta$ of the plane sections, as
shown in Figure \ref{fig:cross_section}(d). From linear theory, it can
be shown that $W(z)$ and $\theta(z)$ may be represented by the
following expressions \cite{HDSM88}:
\begin{equation}
  W(z)=-q_s L \sin \frac {\pi z}{L}, \quad
  \theta(z) = q_t \pi \cos\frac{\pi z}{L},
\end{equation}
where the quantities $q_s$ and $q_t$ are the generalized coordinates of the
sway and tilt components respectively. The corresponding shear strain
$\gamma_{yz}$ during bending is given by the following expression:
\begin{equation}
  \gamma_{yz} = \frac{\D W}{\D z} + \theta = -\left(q_s-q_t\right)\pi
  \cos \frac{\pi z}{L}.
\end{equation} 
In the current model, only geometries are chosen where global buckling
about the $x$-axis is critical.

\begin{figure}[hbt]
\centering
\subfigure[]{\includegraphics[scale=1]{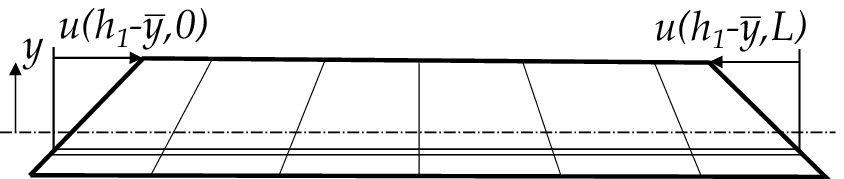}}\quad
\subfigure[]{\includegraphics[scale=1]{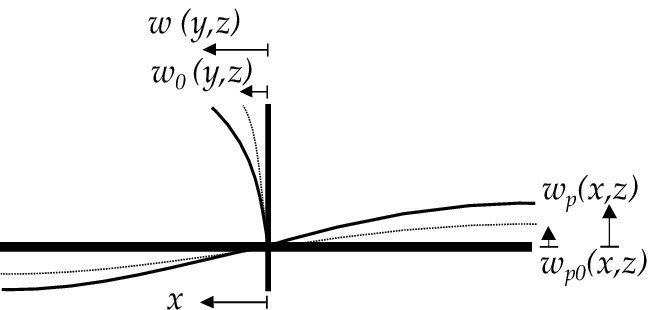}}
\subfigure[]{\includegraphics[scale=1]{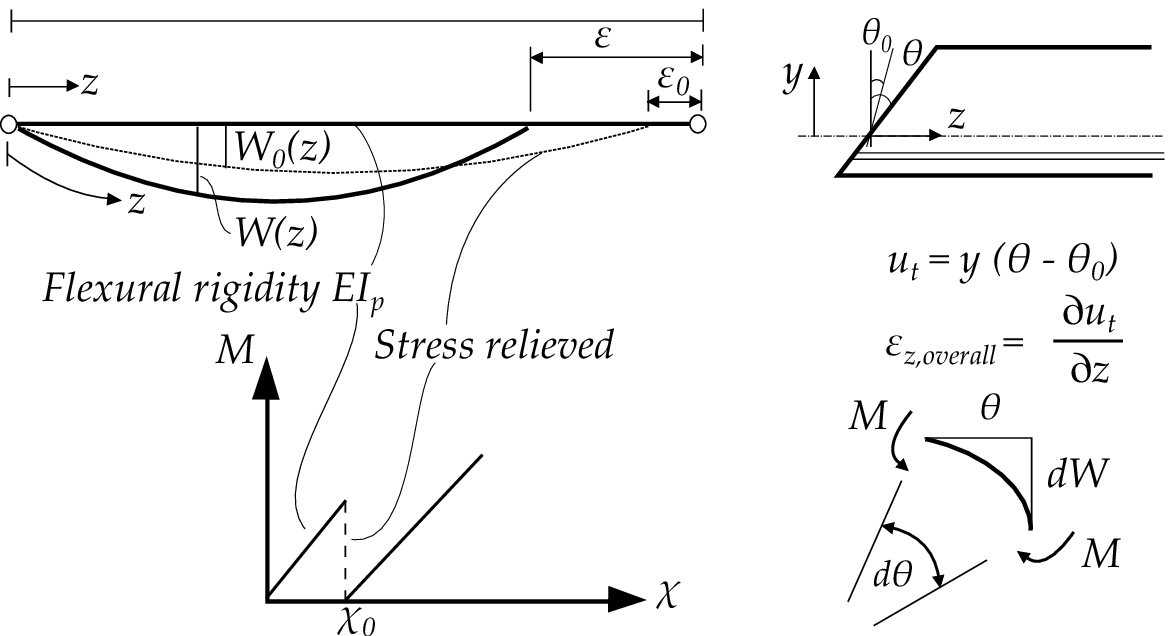}}
\caption{(a) Local in-plane deflection of the stiffener $u(y,z)$. (b)
  Local out-of-plane deflection of the stiffener $w_s(y,z)$ with the
  initial imperfection $w_0(y,z) $ and local out-of-plane deflection
  of the plate $w_p(x,z)$ with the initial imperfection $w_{p0}(x,z)$.
  (c) Introduction of the global imperfection functions $W_0$ and
  $\theta_0$; for the local imperfections, $w_0$ or $w_{p0}$ would
  replace $W_0$ and apply only to the stiffener or the main plate
  respectively.}
\label{fig:local}
\end{figure}

The kinematics of the local buckling modes for the stiffener and the
plate are modelled with appropriate boundary conditions. A linear
distribution in $y$ for the local in-plane displacement $u(y,z)$ is
assumed due to Timoshenko beam theory:
\begin{equation}
  u(y,z) = Y(y) u(z),
\end{equation}
where $Y(y)=(y+\bar{y})/h_1$, as depicted in Figure
\ref{fig:local}(a).

Formulating the assumed deflected shape, however, for out-of-plane
displacements of the stiffener $w(y,z)$ and the main plate $w_p(x,z)$,
see Figure \ref{fig:local}(b), the stiffness of the rotational spring
$c_p$, depicted in Figure \ref{fig:cross_section}(c), is
considered. The role of the spring is to resist the rotational
distortion from the relative bending of the main plate and the
stiffener with respect to the original rigid body configuration.  The
shape of the local buckling mode along the depth of the stiffener and
along the width of the main plate can be therefore estimated, using
the Rayleigh--Ritz method \cite{EnergyMeth}, by a nonlinear function
that is a summation of both polynomial and trigonometric terms. The
general form of these approximations can be expressed by the following
relationships:
\begin{equation}
   w(y,z) = f(y) w(z), \quad
   w_p(x,z) = g(x) w_p(z),
\end{equation}
where:
\begin{equation}
 \begin{aligned}
  f(y) & = B_0 + B_1 Y + B_2 Y^2 + B_3 Y^3 + B_4 \sin \left( \pi Y \right), \\
  g(x) & = C_0 + C_1 X + (-1)^i C_2 X^2 + C_3 X^3
  + C_4\sin\left( \pi X \right),
 \end{aligned} 
\end{equation}
and $X(x)=x/b$. Moreover, for $i=1$, the range $x=[0,b/2]$ and for
$i=2$, the range $x=[-b/2,0]$. For $f(y)$, the constant coefficients
$B_0$, $B_1$, $B_2$, $B_3$ and $B_4$ are determined by applying
appropriate boundary conditions for the stiffener. At the junction
between the stiffener and the main plate, $y=-\bar{y}$, the conditions
are:
\begin{equation}
  w(y,z)=0, \quad
  -D_s\frac{\partial^2}{\partial y^2} w(y,z)=c_p
  \frac{\partial}{\partial y} w(y,z)
  \label{eq:bc:w}
\end{equation}
if the main plate does not rotate, whereas at the stiffener tip,
$y=h_1-\bar{y}$, the conditions are:
\begin{equation}
  D_s\frac{\partial^2}{\partial y^2} w(y,z)=0, \quad
  D_s\frac{\partial^3}{\partial y^3} w(y,z)=0.
  \label{eq:bc:w_tip}
\end{equation}
For $g(x)$, the constant coefficients $C_0$, $C_1$, $C_2$, $C_3$ and
$C_4$ are determined by applying appropriate boundary conditions for
the main plate. At the junction between the stiffener and the main
plate, $x=0$, the conditions are:
\begin{equation}
  w_p(x,z)=0, \quad -D_p\frac{\partial^2}{\partial x^2}
  w_p(x,z)=c_p \frac{\partial}{\partial x} w_p(x,z)
  \label{eq:bc:w_p}
\end{equation}
if the stiffener does not rotate, whereas at the tips of the main
plate at $x=b/2$ and $x=-b/2$, the conditions of which are subtly
different from those given in Equation (\ref{eq:bc:w_tip}) and are
thus:
\begin{equation}
  \frac{\partial}{\partial x}w_p(x,z)=0, \quad
  D_p\frac{\partial^3}{\partial x^3} w_p(x,z)=0,
  \label{eq:bc:w_p_tip}
\end{equation}
where $D_s$ and $D_p$ are the stiffener and the plate flexural
rigidities given by the expressions $Et_s^3/[12(1-\nu^2)]$ and
$Et_p^3/[12(1-\nu^2)]$ respectively. It is worth emphasizing that the
second (mechanical) boundary conditions for determining $B_n$ and
$C_n$, given in Equations (\ref{eq:bc:w}) and (\ref{eq:bc:w_p}), are
simplifying approximations that are admissible since the formulation
is based essentially on the Rayleigh--Ritz method
\cite{EnergyMeth}.

The length $\bar{y}$ gives the location of the neutral-axis of bending
measured from the centre line of the main plate and is expressed thus:
\begin{equation}
  \bar{y} = \frac{t_s\left[ h_1^2 - h_2^2 \right]}{2
    \left[\left(b-t_s\right)t_p + \left(h_1+h_2\right) t_s \right]}.
  \label{eq:ybar}
\end{equation}
The final constants are fixed by imposing the normalizing conditions
with $f(h_1-\bar{y})=1$ and $g(b/2)=1$.  The functions for the
deflected shapes $w(y,z)$ and $w_p(x,z)$ can be written thus:
\begin{equation}
 \begin{aligned}
   w(y,z)&=\left\{ Y - J_s \frac{\pi^3}{6} \left[ 2 Y - 3 Y^2 + Y^3
       - \frac{6}{\pi^3} \sin \left( \pi Y \right)\right] \right\}
   w(z), \\
   w_p(x,z) &= -\left\{ \sin \left( \pi X \right) +J_p \left[ X +
       (-1)^i X^2 -\frac14 \sin \left( \pi X \right) \right] \right\}
   w_p(z),
  \end{aligned}
\end{equation}
where:
\begin{equation}
  J_s = \left\{\pi \left[\frac{D_s \pi^2}{c_p h_1} +
      \frac{\pi^2}{3} - 1\right] \right\}^{-1},\quad
  J_p =\left\{\left[ \frac14 - \frac{2D_p}{c_p b \pi} -
      \frac{1}{\pi} \right] \right\}^{-1}.
\end{equation}
In physical experiments, it is often observed that the main plate
deflects in sympathy with the stiffener to some extent and so in the
current work, the following relationship is assumed, $w_p(z)=\lambda_p
w(z)$.  Since the rotations would therefore be all in the same sense
they can be expressed as first derivatives of $w$ or $w_p$; these are
multiplied by the joint rotational stiffness $c_p$ such that the total
bending moment is established. By allowing both the main plate and the
stiffener to rotate locally and summing the bending moments for the
stiffener and both sides of the main plate at the intersection
($x=0,y=-\bar{y}$), an explicit relationship can be derived:
\begin{equation}
  D_s \frac{\partial^2 w}{\partial y^2} + \left. D_p \frac{\partial^2 
      w_p}{\partial x^2}\right|_{i=1} \left. - D_p \frac{\partial^2 
      w_p}{\partial x^2}\right|_{i=2}= c_p \left( \frac{\partial
      w}{\partial y} + \left. \frac{\partial
        w_p}{\partial x}\right|_{i=1} + \left. \frac{\partial
        w_p}{\partial x}\right|_{i=2} \right).
  \label{eq:jointmoment}
\end{equation}
The negative sign in front of the final term of the left hand side of
Equation (\ref{eq:jointmoment}) reflects the fact the main plate
bending moment changes sign at $x=0$.  The expression for the
deflection relating parameter $\lambda_p$ can be determined by
substituting the aforementioned expression $w_p(z) =\lambda_p w(z)$
into Equation (\ref{eq:jointmoment}) and, after a bit of manipulation,
the following relationship is derived:
\begin{equation}
  \lambda_p=\left(\frac{2 b^2}{3 h_1^2} \right) \left[ \frac{c_p h_1
      \left[ 3 + J_s \pi(3-\pi^2)\right] -3 D_s J_s \pi^3}{%
      8 D_p J_p + c_p b \left[ 4 \pi + J_p (4 - \pi) \right]} \right].
\end{equation}
This simplifies the formulation considerably by allowing the system to
be modelled with effectively only one out-of-plane displacement
function $w$.

\subsection{Imperfection modelling}

\begin{figure}[htb]
\centering
\subfigure[]{\includegraphics[scale=0.7]{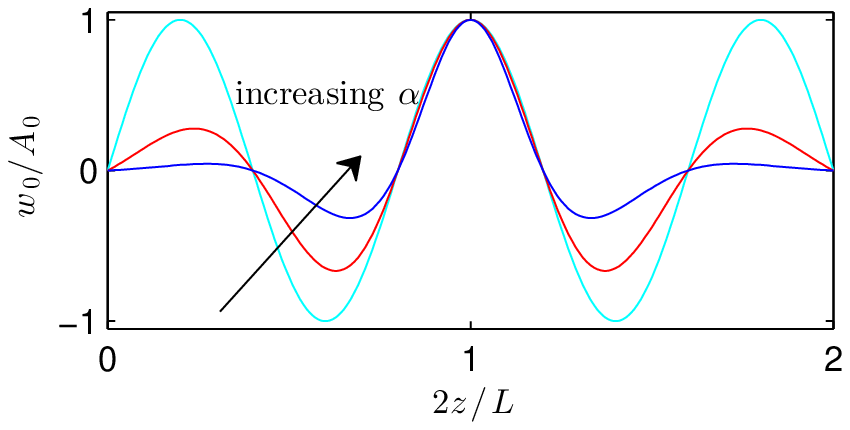}}\qquad
\subfigure[]{\includegraphics[scale=0.7]{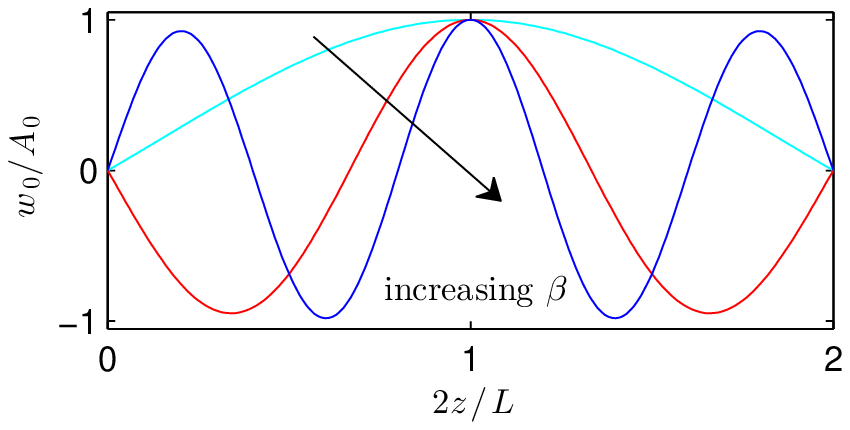}}
\caption{Local imperfection profile $w_0$. (a) Localized imperfections
  introduced by increasing $\alpha$. (b) Periodic imperfections
  ($\alpha=0$) with different numbers of half sine waves by changing
  $\beta$. In both cases $\eta=L/2$.}
\label{fig:impshape}
\end{figure}

Since real structures contain imperfections, the current model
incorporates the possibility of both global and local initial
imperfections within the geometry. This is performed by introducing
initial deflections that are stress-relieved, as shown in Figure
\ref{fig:local}(c), such that the strain energies are zero in the
initially imperfect state. An initial out-of-straightness $W_0$ is
introduced as a global imperfection as well as the corresponding
initial rotation of the plane section $\theta_0$ of the stiffener. The
expressions for these functions are:
\begin{equation}
  W_0(z) = -q_{s0} L \sin\frac{\pi z}{L}, \quad
  \theta_0 = q_{t0} \pi \cos\frac{\pi z}{L},
\end{equation}
with $q_{s0}$ and $q_{t0}$ defining the amplitudes of the global
imperfection. The local out-of-plane imperfection for the stiffener
and the main plate is formulated from a first order approximation of a
multiple scale perturbation analysis of a strut on a softening elastic
foundation \cite{WHW97}, the mathematical shape of which is expressed
as:
\begin{equation}
  w_0(z) = A_0 \sech \left[\frac{\alpha\left(z-\eta\right)}{L}\right] 
  \cos \left[ \frac{\beta \pi \left(z-\eta \right)}{L}\right],
  \label{eq:localimp}
\end{equation}
where $z = [0,L]$ and $w_0$ is symmetric about $z=\eta$. This
function has been shown in the literature to provide a representative
imperfection for local--global mode interaction problems
\cite{Wadee2000}. Moreover, this form for $w_0$ enables the study
of periodic and localized imperfections; a local imperfection is
periodic when $\alpha=0$ with a number of half sine waves equal to
$\beta$ along the length of the panel. It is also noted that the
relationship between $w_{p0}$ and $w_0$ corresponds to that for the
perfect case; hence, $w_{p0}=\lambda_p w_0$ is assumed. The shape
of the initial imperfection is illustrated in Figure
\ref{fig:impshape}.
By increasing the $\alpha$ value, the initial imperfection forms into
a localized shape, as shown in Figure \ref{fig:impshape}(a), otherwise
the imperfection shape is periodic, as shown in Figure
\ref{fig:impshape}(b).

\subsection{Total potential energy}

A well established procedure for deriving the total potential energy
$V$, has been presented in previous work \cite{WF1_2014}; the current
work follows the same approach but now includes the local buckling
deflection of the main plate.  The global strain energy $U_{bo}$ due
to Euler buckling is given by the equation below:
\begin{equation}
\begin{aligned}
  U_{bo} & = \frac12 EI_p \int_0^L \left( \ddot{W}-\ddot{W}_0
  \right)^2 \D z=\frac12 EI_p \int_0^L \left(q_s-q_{s0}\right)^2
  \frac{\pi^4}{L^2}\sin^2\frac{\pi z}{L}\D z,
\end{aligned}
\label{eq:ob}
\end{equation}
\noindent
where dots represent differentiation with respect to $z$ and
$I_p=(b-t_s)t_p^3/12+(b-t_s)t_p \bar{y}^2$ is the second moment of
area of the plate about the global $x$-axis.  The strain energy from
local bending of the stiffener and the main plate $U_{bl}$ is given by
the following expression:
\begin{equation}
\begin{split}
  U_{bl} & = \frac{D_s}{2} \int_0^L \int_{-\bar{y}}^{h_1-\bar{y}}
  \left\{ \left[ \frac{\partial^2(w-w_0)}{\partial
        z^2}+\frac{\partial^2(w-w_0)}{\partial y^2} \right]^2 \right. \\
  & \left.  \qquad - 2\left(1-\nu \right) \left[\frac{\partial^2
        (w-w_0)}{\partial z^2}
      \frac{\partial^2(w-w_0)}{\partial
        y^2}-\left(\frac{\partial^2(w-w_0)}{\partial z \partial
          y}
      \right)^2 \right] \right\} \D y \D z \\
  & \quad + \frac{D_p}{2} \int_0^L \int_{-b/2}^{b/2} \left\{ \left[
      \frac{\partial^2(w_p-w_{p0})}{\partial
        z^2}+\frac{\partial^2(w_p-w_{p0})}{\partial x^2} \right]^2
  \right. \\
  & \left.  \qquad - 2\left(1-\nu \right) \left[\frac{\partial^2
        (w_p-w_{p0})}{\partial z^2}
      \frac{\partial^2(w_p-w_{p0})}{\partial
        x^2}-\left(\frac{\partial^2(w_p-w_{p0})}{\partial z \partial
          x} \right)^2 \right] \right\} \D x \D z, \\
  \\
  & = \frac{D_s}{2} \int_0^L \left[\{ f^2 \}_y \left(\ddot{w}-
      \ddot{w}_0 \right)^2+\bigl\{ f''^2 \bigr\}_y \left(w-w_0
    \right)^2 +2 \nu \bigl\{ f f'' \bigr\}_y
    (w-w_0)(\ddot{w}-\ddot{w}_0) \right. \\
  & \left.  \qquad +2(1-\nu) \bigl\{ f'^2 \bigr\}_y
    (\dot{w}-\dot{w}_0)^2
  \right] \D z + \frac{D_p}{2} \int_0^L \left[\{ g^2 \}_x \left(\ddot{w}_p-
      \ddot{w}_{p0} \right)^2 + \bigl\{ g''^2 \bigr\}_x \left(w_p-w_{p0}
    \right)^2 \right. \\
    & \qquad \left. + 2 \nu \bigl\{ g g'' \bigr\}_x
    (w_p-w_{p0})(\ddot{w}_p-\ddot{w}_{p0}) + 2(1-\nu) \bigl\{ g'^2 \bigr\}_x
    (\dot{w}_p-\dot{w}_{p0})^2 \right] \D z.
\end{split}
\end{equation}
The terms within the braces are definite integrals, thus:
\begin{equation}
\{F(y)\}_y=\int_{-\bar{y}}^{h_1-\bar{y}} F(y) \D y, \qquad
\{H(x)\}_x=\int_{-b/2}^{b/2} H(x) \D x,
\end{equation}
where $F$ and $H$ are example functions representing the actual
expressions within the braces and primes denote differentiation with
respect to the subscript outside the closing brace.

The membrane energy $U_m$ is derived from the direct strains
($\varepsilon$) and shear strains ($\gamma$) in the plate and the
stiffener. It is thus:
\begin{equation}
\begin{aligned}
  U_m & = U_d + U_s \\
  & = \frac12 \int_0^L \left\{ \int_{-t_s/2}^{t_s/2} \biggl[
    \int_{-\bar{y}}^{h_1-\bar{y}} \left( E \varepsilon_{zt}^2 + G
      \gamma_{yzt}^2 \right)\D y + \int_{-(h_2+\bar{y})}^{-\bar{y}}
    \left( E \varepsilon_{zb}^2
      + G \gamma_{yzb}^2 \right) \D y \biggr] \D x \right. \\
  & \qquad \left. + \int_{-t_p/2}^{t_p/2}
    \int_{-(b-t_s)/2}^{(b-t_s)/2} E \varepsilon_{zp}^2 \D x \D y
  \right\} \D z.
\end{aligned}
\end{equation}
Note that the transverse component of the strain $\varepsilon_y$ is
neglected since it has been shown that it has no effect on the
post-buckling behaviour of a long plate with three simply-supported
edges and one free edge \cite{Koiter1976}. The global buckling
contribution for the longitudinal strain $\varepsilon_z$ can be
obtained from the tilt component of the global mode, which is given
by:
\begin{equation}
  \varepsilon_{z,\mathrm{global}} = y\frac{\partial \theta}{\partial z}
  = -y\left(q_t-q_{t0}\right)\frac{\pi^2}{L}\sin\frac{\pi z}{L}.
\end{equation}
The local mode contribution is based on von K\'arm\'an plate theory
\cite{Bulson}. A pure in-plane compressive strain $\Delta$ is also
included. The combined expressions for the direct strains for the top
and bottom stiffeners $\varepsilon_{zt}$ and $\varepsilon_{zb}$
respectively, and for the main plate $\varepsilon_{zp}$ are given
thus:
\begin{equation}
\begin{aligned}
  \varepsilon_{zt} & = -y \left(q_t-q_{t0}\right)\frac{\pi^2}{L} \sin
  \frac{\pi z}{L}-\Delta + \frac{\partial u}{\partial z} + \frac12
  \left(\frac{\partial w}{\partial z} \right)^2 - \frac12
  \left(\frac{\partial w_0}{\partial z} \right)^2, \\
  & = -y \left(q_t-q_{t0}\right)\frac{\pi^2}{L} \sin \frac{\pi z}{L} -
  \Delta + Y \dot{u} + \frac12 \{f^2\}_y
  \left(\dot{w}^2-\dot{w}_0^2 \right),\\
  \varepsilon_{zb} & = -y \left(q_t-q_{t0}\right)\frac{\pi^2}{L} \sin
  \frac{\pi z}{L}-\Delta \\
\varepsilon_{zp} & = -\Delta + \frac12
  \left(\frac{\partial w_p}{\partial z} \right)^2 - \frac12
  \left(\frac{\partial w_{p0}}{\partial z} \right)^2.
  \end{aligned} 
\end{equation}
The membrane energy contribution from the direct strains $U_d$ is
therefore:
\begin{equation}
\begin{aligned}
  U_d = \frac12 E t_s & \int_0^L  \biggl\{ \frac13\left[
    \left(h_1-\bar{y} \right)^3 + \left(h_2+\bar{y} \right)^3 \right]
  \left(q_t-q_{t0}\right)^2 \frac{\pi^4}{L^2} \sin^2 \frac{\pi z}{L} +
  \Delta^2 \left(h_1+h_2 \right) \\
  & + \left[ \left(h_1-\bar{y}
    \right)^2 - \left(h_2+\bar{y} \right)^2 \right] \Delta
  \left(q_t-q_{t0} \right) \frac{\pi^2}{L} \sin \frac{\pi z}{L} \\
  & + h_1 \left[\frac13 {\dot{u}}^2 + \frac{1}{4h_1}\{f^4\}_y
    \left(\dot{w}^2-\dot{w}_0^2 \right)^2 +
    \biggl\{\frac{Yf^2}{h_1}\biggr\}_y \dot{u}
    \left(\dot{w}^2-\dot{w}_0^2 \right)\right] \\
  & - \left(q_t-q_{t0}\right) \frac{h_1 \pi^2}{L} \sin \frac{\pi z}{L}
  \left[ \left(\frac23 h_1-\bar{y} \right) \dot{u}
   + \frac{1}{h_1} \{yf^2\}_y \left( \dot{w}^2-\dot{w}_0^2 \right)
  \right]  \\
  & - h_1 \Delta \left[ \dot{u} + \frac{1}{h_1}\{f^2\}_y
    \left(\dot{w}^2-\dot{w}_0^2 \right) \right] \\
 & + \left(\frac{t_p}{t_s}\right)\left[ (b-t_s) \Delta^2 + \frac14
   \{g^4\}_x \left( \dot{w}_p^2-\dot{w}_{p0}^2 \right)^2-\Delta
   \{g^2\}_x \left( \dot{w}_p^2-\dot{w}_{p0}^2 \right) \right]
 \biggr\}
  \D z.
\end{aligned}
\end{equation}
The membrane energy contribution from shear strains arises from those
in the main plate $\gamma_{xz}$ as well those in the stiffeners
$\gamma_{yz}$; the respective general expressions being:
\begin{equation}
\begin{aligned}
  \gamma_{xz} & = \frac{\partial w_p}{\partial z}\frac{\partial
    w_p}{\partial x} - \frac{\partial w_{p0}}{\partial
    z}\frac{\partial w_{p0}}{\partial x},\\
  \gamma_{yz} & = \frac{\partial}{\partial z} \left(W - W_0\right)
  + \left(\theta-\theta_0\right) + \frac{\partial u}{\partial y} +
  \frac{\partial w}{\partial z}\frac{\partial w}{\partial y} -
  \frac{\partial w_0}{\partial z}\frac{\partial w_0}{\partial y},
\end{aligned}
\end{equation}
hence, the expressions for the top and bottom stiffeners and the plate
are given respectively:
\begin{equation}
\begin{aligned}
  \gamma_{yzt} & = -\left[
    \left(q_s-q_{s0}\right)-\left(q_t-q_{t0}\right)\right] \pi \cos
  \frac{\pi z}{L} +\frac{u}{h_1} + \{ff'\}_y
  (w\dot{w}-w_0\dot{w}_0),\\ 
  \gamma_{yzb} & =-\left[ \left(q_s-q_{s0}\right) - \left(q_t - q_{t0}
    \right)\right] \pi \cos \frac{\pi z}{L},
\end{aligned}
\end{equation}
with the explicit expression for the main plate shear strain:
\begin{equation}
  \gamma_{xz} = \{gg'\}_x (w_p\dot{w}_p-w_{p0}\dot{w}_{p0}).  
\end{equation}
The membrane energy contribution from the shear strains $U_s$ is
therefore:
\begin{equation}
\begin{aligned}
  U_s = \frac12 G t_s & \int_0^L \biggl\{ \left[ \left(q_s-q_{s0}
    \right)-\left(q_t-q_{t0} \right)\right]^2 {\pi}^2 \cos^2 \frac{\pi
    z}{L} \left(h_1+h_2 \right)\\
  & + \frac{1}{h_1} \left[ u^2 + h_1 \{\left(ff'\right)^2\}_y 
   \left( w\dot{w}-w_0\dot{w}_0 \right)^2+ 2 \{ f f'\}_y u
   \left( w\dot{w}-w_0\dot{w}_0 \right) \right] \\
  & - \left[ \left(q_s-q_{s0} \right)-\left(q_t-q_{t0} \right)\right]
  \biggl[ 2u+2\{ff'\}_y (w\dot{w}-w_0\dot{w}_0 )\biggr] \pi
  \cos \frac{\pi z}{L} \\ 
 & + \left(\frac{t_p}{t_s}\right) \bigl\{\left(g g'\right)^2
 \bigr\}_x (w_p\dot{w}_p-w_{p0} \dot{w}_{p0})^2 \biggr\} \D z.
\end{aligned}
\end{equation}

The final component of strain energy is that stored in the rotational
spring of stiffness $c_p$ representing the rigidity of the main
plate--stiffener joint. It is given thus:
\begin{equation}
\begin{aligned}
  U_{sp} & = \frac12 c_p \int_0^L \biggl\{ \biggl[
  \frac{\partial}{\partial y} \left[ w(-\bar{y},z)-w_0(-\bar{y},z)
  \right] - \frac{\partial}{\partial x} \left[ w_p(0,z)-w_{p0}(0,z)
  \right] \biggr]^2 \biggr\} \D z, \\
 & = \frac12 c_p \int_0^L \biggl\{ \biggl[  f'(-\bar{y}) \left(
   w-w_0 \right) - g'(0) (w_p-w_{p0})\biggr]^2 \biggr\}  \D z, 
  \end{aligned}
\end{equation}
where $f'(-\bar{y})$ and $g'(0)$ indicate the values of $f'$ and $g'$
at $y=-\bar{y}$ (or $Y=0$) and $x=0$ respectively. The final component
of $V$ is the work done by the axial load $P$, which is given by:
\begin{equation}
  P\mathcal{E} = \frac{P}{2} \int_0^L \left[ 2\Delta + q_s^2 \pi^2
    \cos^2 \frac{\pi z}{L} - 2\left( \frac{h_2+\bar{y}}{h_1+h_2}
    \right) \dot{u} \right] \Dx z,
  \label{eq:WD}
\end{equation}
where the end-displacement $\mathcal{E}$ comprises components from
pure squash, sway from global buckling and the component from local
buckling of the stiffener. Therefore, the total potential energy $V$
is given by the summation of all the strain energy terms minus the
work done, thus:
\begin{equation}
   V = U_{bo} + U_{bl} + U_m +U_{sp}- P\mathcal{E}.
\end{equation}

\subsection{Variational Formulation}

The governing equations of equilibrium are obtained by performing the
calculus of variations on the total potential energy $V$ following the
well established procedure presented in previous work
\cite{HW1998,WF1_2014}. The integrand of the total potential energy
$V$ can be expressed as the Lagrangian ($\mathcal{L}$) of the form:
\begin{equation}
  V=\int_0^L \mathcal{L}\left(\ddot w, \dot{w}, w, \dot {u}, u,
    z \right) \D z,
\end{equation}
of course, this is after substituting the relationship, $w_p =
\lambda_p w$. Hence, the first variation of $V$ is:
\begin{equation}
  \delta V = \int_0^L \left[ \frac{\partial \mathcal{L}}{\partial
      \ddot w} \delta \ddot w + \frac{\partial
      \mathcal{L}}{\partial \dot w} \delta \dot w + \frac{\partial
      \mathcal{L}}{\partial w} \delta w
    + \frac{\partial \mathcal{L}}{\partial \dot u}
    \delta \dot u + \frac{\partial \mathcal{L}}{\partial u} \delta u
  \right] \D z.
\end{equation}
To determine the equilibrium states, $V$ must be stationary, hence the
first variation $\delta V$ must vanish for any small change in $w$ and
$u$. Since $\delta\ddot w=\Dx(\delta\dot w)/\D z$, $\delta\dot w=\Dx
(\delta w)/\Dx z$ and similarly $\delta \dot u = \Dx (\delta u)/\Dx
z$, integration by parts allows the development of the Euler--Lagrange
equations for $w$ and $u$; these comprise a fourth-order and a
second-order nonlinear differential equation for $w$ and $u$
respectively. To facilitate the solution within the package
\textsc{Auto-07p}, the variables are rescaled with respect to the
non-dimensional spatial coordinate $\tilde z$, defined as $\tilde z =
2z/L$. Similarly, non-dimensional out-of-plane and in-plane
displacements $\tilde{w}$ and $\tilde{u}$ are defined with the
scalings $2w/L$ and $2u/L$ respectively. Note that the scalings
exploit symmetry about midspan and the equations are hence solved for
half the strut length; this assumption has been shown to be perfectly
acceptable for cases where global buckling is critical
\cite{Wadee2000}. The non-dimensional differential equations for $w$
and $u$ are thus:
\begin{equation}
  \begin{aligned}
    & \left[ 1 + \lambda_p^2 \left(\frac{t_p}{t_s}\right)^3
      \frac{\bigl\{g^2\bigr\}_x}{\bigl\{f^2\bigr\}_y} \right] \left(
      \tilde{\ddddot{w}}-\tilde{\ddddot{w_0}} \right)+
    \frac{L^2}{2\{f^2\}_y} \biggl\{ \left[ \nu \{f f''\}_y - (1-\nu)
      \{ f'^2
      \}_y \right] \\
    & \qquad \quad + \lambda_p^2 \left(\frac{t_p}{t_s}\right)^3
    \left[\nu \{g g''\}_x - (1-\nu) \{g'^2\}_x \right]\biggr\}\left(
      \tilde{\ddot{w}} - \tilde{\ddot{w}}_0\right) + \tilde{k}
    \left(\tilde{w}-\tilde{w_0}\right)\\
    & \qquad - \tilde{D} \biggl[ \frac{\{ f^4\}_y}{\{f^2\}_y} \left( 3
      \tilde{\dot{w}}^2 \tilde{\ddot{w}} - \tilde{\ddot{w}}
     \tilde{\dot{w}}_0^2 - 2 \tilde{\ddot{w}}_0\tilde{\dot{w}}_0
      \tilde{\dot{w}} \right) + \frac{ \{2 Y f^2 \}_y}{\{f^2\}_y}
    \left( \tilde{\ddot{u}}
      \tilde{\dot{w}} + \tilde{\ddot{w}} \tilde{\dot{u}} \right) \\
    & \qquad \quad - 2 \Delta \tilde{\ddot{w}} - 2 \left( q_t-q_{t0} \right)
    \frac{\pi^2}{L} \frac{\{ yf^2 \}_y}{\{f^2\}_y} \left( \sin
      \frac{\pi \tilde{z}}{2} \tilde{\ddot{w}} + \frac{\pi}{2} \cos
      \frac{\pi \tilde{z}}{2} \tilde{\dot{w}}
    \right) \biggr] \\
    & \qquad - \frac{ \tilde{G} L^2 \tilde{w}}{2 \{f^2\}_y} \biggl[ \{
    \left( f f' \right)^2 \}_y \left( \tilde{\dot{w}}^2 + \tilde{w}
      \tilde{\ddot{w}} - \tilde{\dot{w}}_0^2 - \tilde{w}_0
      \tilde{\ddot{w}}_0 \right) + \frac1 h_1 \{f f'\}_y
    \tilde{\dot{u}}  \\
    & \qquad \quad + \left[ \left( q_s-q_{s0} \right)- \left( q_t
        -q_{t0} \right) \right] \frac{\pi^2}{L} \{f f'\}_y \sin
    \frac{\pi
      \tilde{z}}{2} \biggr] \\
    & \qquad - \left( \frac{t_p}{t_s} \right) \frac{
      \tilde{D}\lambda_p^2}{\{f^2\}_y} \biggl[ \lambda_p^2 \{ g^4 \}_x
    \left( 3 \tilde{\dot{w}}^2 \tilde{\ddot{w}} - \tilde{\ddot{w}}
      {\dot{w}}_0^2 - 2 \tilde{\ddot{w}}_0 \tilde{\dot{w}}_0
      \tilde{\dot{w}} \right)
    - 2 \Delta \{g^2 \}_x \tilde{\ddot{w}} \biggr] \\
    & \qquad - \left( \frac{t_p}{t_s} \right) \frac{L^2 \tilde{G} \lambda_p^4
      \tilde{w}}{2 \{f^2\}_y} \biggl[ \{ \left( g g' \right)^2 \}_x
    \left( \tilde{\dot{w}}^2 + \tilde{w} \tilde{\ddot{w}} -
      \tilde{\dot{w}}_0^2 - \tilde{w}_0 \tilde{\ddot{w}}_0 \right)
    \biggr] = 0,
\end{aligned}
\label{eq:wdddd}
\end{equation}
\begin{equation}
\begin{aligned}
  \tilde{\ddot{u}} & - \frac34 \frac{\tilde{G}}{\tilde{D}} \psi
  \biggl[ \psi \left( \tilde{u}+ \{ f f'\}_y \left( \tilde{w}
      \tilde{\dot{w}} - \tilde{w}_0 \tilde{\dot{w}}_0
    \right)\right) - 2 \pi \left[
    \left(q_s-q_{s0}\right)-\left(q_t-q_{t0}\right) \right] \cos
  \frac{\pi \tilde{z}}{2} \biggr] \\
  & -\left\{\frac{3 Y f^2}{h_1} \right\}_y \left(
    \tilde{\dot{w}}\tilde{\ddot{w}} + \tilde{\dot{w}}_0
    \tilde{\ddot{w}}_0 \right) + \frac12 \left(q_t-q_{t0}\right)
  \pi^3 \left( \psi - \frac{3\bar{y}}{2L} \right) \cos \frac{\pi
    \tilde{z}}{2} = 0,
\end{aligned}
\label{eq:udd}
\end{equation}
where the non-dimensional parameters are defined thus:
\begin{equation}
\begin{aligned}
  \tilde{D}&=\frac{Et_sL^2}{8D_s}, \quad \tilde{G}=\frac{Gt_sL^2}{8D_s}, \\
  \tilde{k}&= \frac{L^4}{16 \left\{f^2 \right\}_y} \biggl[
  \left\{f''^2\right\}_y + \lambda_p^2
  \left(t_p/t_s\right)^3\{g''^2\}_x + c_p \left[
    f'(-\bar{y})-\lambda_p g'(0)\right]^2/D_s \biggr],
  \end{aligned}
\end{equation}
and $\tilde{w}_0 = 2 w_0/L$, $\psi=L/h_1$. There are further
equilibrium conditions that relate to $V$ being minimized with respect
to the generalized coordinates $q_s$, $q_t$ and $\Delta$.  This leads
to the derivation of three integral conditions in non-dimensional
form as follows:
\begin{equation}
  \begin{split}    
  \frac{\partial V}{\partial q_s} & = \pi^2 \left(q_s - q_{s0}\right) +
  \tilde{s} \left[ \left(q_s - q_{s0}\right)-\left(q_t-q_{t0}\right)
  \right] -\frac{PL^2}{EI_p} q_s \\
  & \quad -\frac{\tilde{s}\tilde{\phi}}{2\pi}\int_0^2 \cos
  \frac{\pi\tilde{z}}{2} \left[ \tilde{u}+ \left\{f f' \right\}_y
    \left( \tilde{w} \tilde{\dot{w}} - \tilde{w}_{0}
      \tilde{\dot{w}}_{0} \right) \right] \Dx \tilde{z} =0,
  \end{split}
\label{eq:equil_int_qs}
\end{equation}
\begin{equation}
  \begin{split}
  \frac{\partial V}{\partial q_t} & = \pi^2 \left(q_t-q_{t0}\right) +
  \tilde{\Gamma_3} \Delta -\tilde{t} \left[
    \left(q_s-q_{s0}\right)-\left(q_t-q_{t0}\right) \right] - \frac12
  \int_0^2 \biggl\{ \sin \frac{\pi \tilde{z}}{2} \left[
    \tilde{\Gamma_1}\tilde{\dot{u}} \right.\\
  & \quad \left.+ \tilde{\Gamma_2} \left( \tilde{\dot{w}}^2
      - \tilde{\dot{w}}_0^2 \right) \right]
  + \frac{\tilde{t} \tilde{\phi}}{\pi} \cos \frac{\pi \tilde{z}}{L} \left[
    \tilde{u}+ \left\{ f f' \right\}_y
    \left(\tilde{w}\tilde{\dot{w}} - \tilde{w_{0}}
    \tilde{\dot{w}}_{0} \right) \right]\biggr\}
  \Dx \tilde{z} = 0,
  \end{split}  
  \label{eq:equil_int_qt}
\end{equation}
\begin{equation}
  \begin{split}
  \frac{\partial V}{\partial \Delta} & = \Delta \left[1+\frac{h_2}{h_1}+
    \frac{t_p(b-t_s)}{t_s h_1} \right] -\frac{P}{Et_sh_1} +
  \left(q_t-q_{t0} \right) \frac{\pi}{Lh_1} \left[\left(h_1-\bar{y} \right)^2 -
      \left(h_2+\bar{y} \right)^2 \right] \\
  & \quad -\frac{1}{4} \int_0^2 \left[ \tilde{\dot{u}} +\frac1 h_1 \left\{
      f^2 \right\}_y
    \left(\tilde{\dot{w}}^2-\dot{w}_0^2 \right) \right] \Dx
  \tilde{z} - \left(\frac{t_p}{t_s} \right) \frac{\lambda_p^2}{4h_1}
  \int_0^2 \biggl[\{g^2\}_x \left(\tilde{\dot{w}}^2-\dot{w}_0^2
  \right) \biggr] \Dx \tilde{z} =0.
\end{split}
\label{eq:equil_int}
\end{equation}
where the rescaled quantities are defined thus:
\begin{equation}
\begin{aligned}
  \tilde{\Gamma_1} &=
  \frac{Lh_1\left(2h_1-3\bar{y}\right)}{\left(h_1-\bar{y}\right)^3 +
    \left(h_2+\bar{y}\right)^3},
  \quad \tilde{\Gamma_2}=\frac{3L\left\{ yf^2\right\}_y}
  {\left(h_1-\bar{y}\right)^3+\left(h_2+\bar{y}\right)^3}, \\
  \tilde{\Gamma_3} &=\frac{6L \left[ \left(h_1-\bar{y}\right)^2
      -\left(h_2+\bar{y} \right)^2 \right]}{\pi \left[
      \left(h_1-\bar{y}\right)^3 +\left(h_2+\bar{y} \right)^3
    \right]}, \quad \tilde{\phi} = \frac{L}{h_1+h_2}, \\
  \tilde{s} &= \frac{G t_s (h_1+h_2) L^2}{EI_p},\quad
  \tilde{t} =\frac{3GL^2(h_1+h_2)}{E\left[\left(h_1-\bar{y}\right)^3
      +\left(h_2+\bar{y} \right)^3 \right]}.
\end{aligned}
\end{equation}
Since the stiffened panel is an integral member, Equation
(\ref{eq:equil_int}) provides a relationship linking $q_s$ and $q_t$
before any interactive buckling occurs, \emph{i.e.}\ when
$w=u=0$. This relationship is also assumed to hold between $q_{s0}$
and $q_{t0}$, which has the beneficial effect of reducing the number
of imperfection amplitude parameters to one; this relationship is
given by:
\begin{equation}
q_{s0}=\left(1+\pi^2 /\tilde{t} \right)q_{t0}.
\end{equation}
The boundary conditions for $\tilde{w}$ and $\tilde{u}$ and their
derivatives are for pinned conditions for $\tilde{z}=0$ and for
reflective symmetry at $\tilde{z}=1$:
\begin{equation}
\label{eq:bc_w}
\tilde{w}(0) = \tilde{\ddot{w}}(0) = \tilde{\dot{w}}(1) =
\tilde{\dddot{w}}(1) = \tilde{u}(1) = 0,
\end{equation}
with a further condition from matching the in-plane strain:
\begin{equation}
  \label{eq:bc_ud}
  \frac13 \tilde{\dot{u}}(0) + \frac12 \left\{ \frac{Y}{h_1} f^2
  \right\}_y \left[ \tilde{\dot{w}}^2(0)-\dot{w}_0^2(0) \right] -
  \frac12 \Delta + \frac{P}{Et_sh_1} \left(
    \frac{h_2+\bar{y}}{h_1+h_2}\right) = 0.
\end{equation}
Linear eigenvalue analysis for the perfect column ($q_{s0}=q_{t0}=0$)
is conducted to determine the critical load for global buckling
$\Pco$. This is achieved by considering the Hessian matrix $\mathbf{V}_{ij}$,
thus:
\begin{equation}
\mathbf{V}_{ij}=\left[
\begin{array}{cc}
  \frac{\partial^2 V}{\partial q_s^2} & \frac{\partial^2
    V}{\partial q_s \partial q_t} \\
  \frac{\partial^2 V}{\partial q_t \partial q_s} & \frac{\partial^2
    V}{\partial q_t^2}
\end{array}
\right],
\end{equation}
where the matrix $\mathbf{V}_{ij}$ is singular at the global critical
load $\Pco$. Hence, the critical load for global buckling is:
\begin{equation}
  \Pco = \frac{\pi^2 EI_p}{L^2} \left[1+ \frac{\tilde{s}}{
      \pi^2+ \tilde{t}} \right].
  \label{eq:pc}
\end{equation}
If the limit $G \rightarrow \infty$ is taken, which represents a
principal assumption in Euler--Bernoulli bending theory, it can be
shown that the critical load expression converges to the Euler
buckling load for the modelled strut, as would be expected.

\section{Numerical results}
\label{sec:numerical}

Numerical results with a varying rotational spring stiffness $c_p$ are
now presented for the perfect system. The continuation and bifurcation
software \textsc{Auto-07p} \cite{auto} is used to solve the complete
system of equilibrium equations presented in the previous section. An
example set of section and material properties are chosen thus:
$L=5000~\mm$, $b=120~\mm$, $t_p = 2.4~\mm$, $t_s=1.2~\mm$,
$h_1=38~\mm$, $h_2=1.2~\mm$, $E = 210~\mathrm{kN/mm^2}$, $\nu = 0.3$.
The global critical load $\Pco$ can be calculated using Equation
(\ref{eq:pc}), whereas the local buckling critical stress
$\sigma_l^\mathrm{C}$ can be evaluated using the well-known formula
$\sigma_l^\mathrm{C}=k_p D\pi^2/(b^2t)$, where the coefficient $k_p$
depends on the plate boundary conditions. By increasing the $c_p$
value, the relative rigidity between the main plate and the stiffener
varies from being completely pinned ($c_p=0$) to fully-fixed ($c_p
\rightarrow \infty$). Therefore the limiting values for $k_p$ are
$0.426$ or $1.247$ for a long stiffener connected to the main plate
with one edge free and the edge defining the junction between the
stiffener and the main plate being taken to be pinned or fixed
respectively \cite{Bulson}. However, the value of the global critical
buckling load $\Pco$ remains the same since it is independent of
$c_p$.

\begin{figure}[hbt]
\centering
\subfigure[]{\includegraphics[scale=0.80]{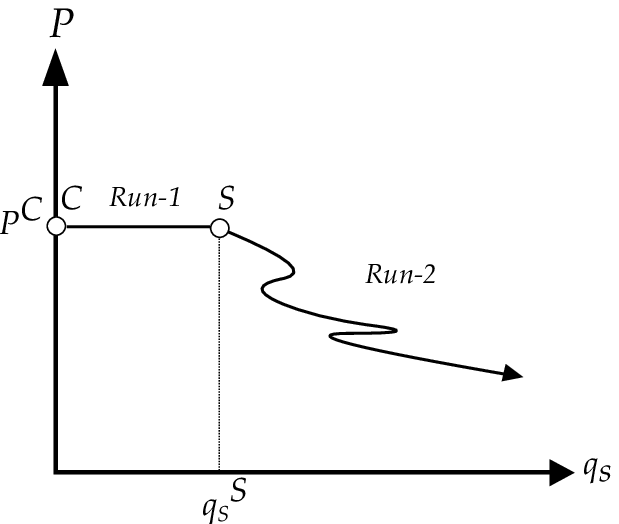}}\quad
\subfigure[]{\includegraphics[scale=0.80]{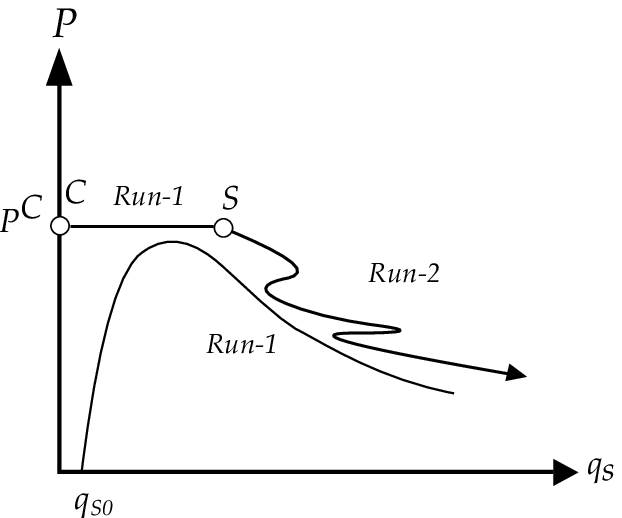}}\quad
\subfigure[]{\includegraphics[scale=0.80]{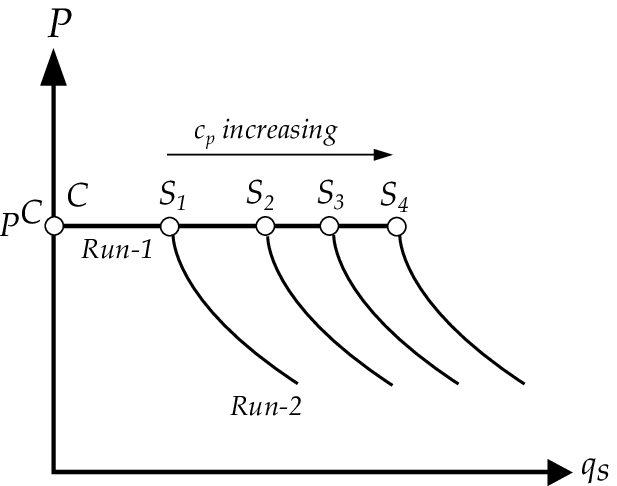}}
\caption{Diagrammatic representation of the sequence of computing the
  equilibrium paths for the (a) perfect and (b) imperfect cases; (c)
  shows the perfect cases for different values of $c_p$ with the
  corresponding secondary bifurcation points changing as $c_p$ is
  increased.}
\label{fig:path}
\end{figure}

To find the equilibrium path in the fundamental and post-buckling
states, a similar solution strategy is performed as in recent work
\cite{WB2014,WF1_2014}, which is illustrated diagrammatically in
Figure \ref{fig:path}.
For a perfect strut, the initial post-buckling path is computed first
from the critical buckling load $\Pc(\equiv\Pco)$ with $q_s$ being
varied. Many bifurcation points are detected on the weakly stable
post-buckling path; the focus being on the one with the lowest value
of $q_s$, termed the secondary bifurcation point $\mathrm{S}$, see
Figure \ref{fig:path}(a). Note that the corresponding $q_s$ value is
labelled as $q_s^\mathrm{S}$. For an imperfect strut, however, the
equilibrium path is computed initially from zero axial load and then
$P$ is increased up to the maximum value where a limit point is
detected. The load subsequently drops and the path is asymptotic to
the perfect path, as shown in Figure \ref{fig:path}(b). If the joint
stiffness $c_p$ is varied, the value of $q_s^\mathrm{S}$ would be
expected to increase, see Figure \ref{fig:path}. This is owing to the
local buckling critical stress increasing, which in turn causes the
required global mode amplitude to trigger local buckling to increase
also.

It is worth noting that for the perfect case, the model is in fact
only valid where global buckling or stiffener local buckling is
critical since the assumption is made such that the main plate can
only buckle in sympathy with the stiffener. For the stiffener local
buckling being critical, the bifurcation would occur when
$P=P_l^\mathrm{C}$ and a stable post-buckling path would initially
emerge from the fundamental path. To include the main plate buckling
locally first, the explicit link between $w_p$ and $w$ would have to
be broken.

\begin{figure}[hbt]
\centering
\subfigure[]{\includegraphics[scale=0.75]{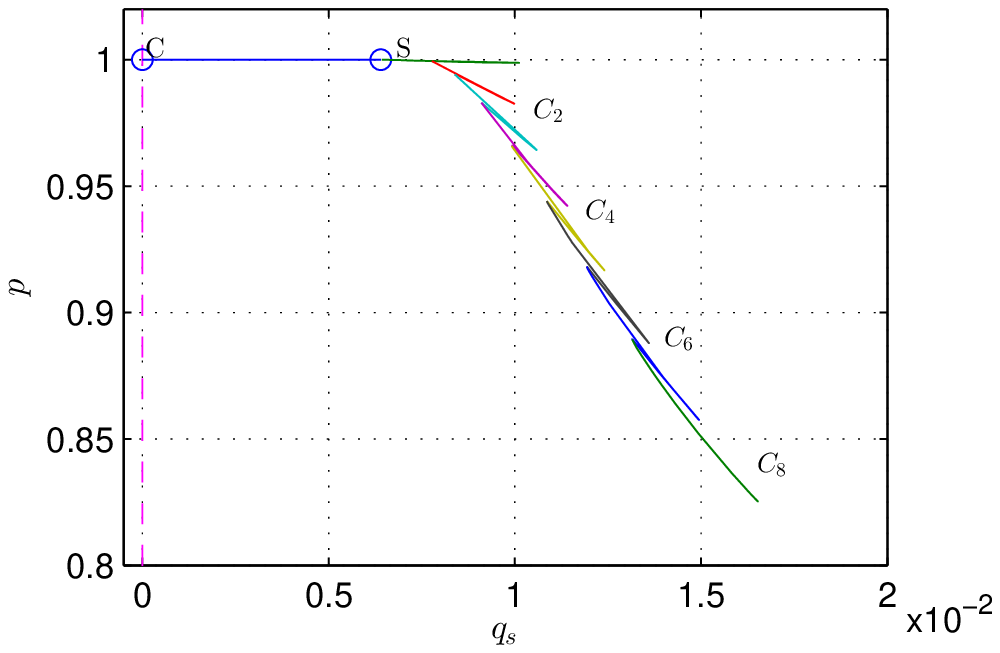}}\quad
\subfigure[]{\includegraphics[scale=0.75]{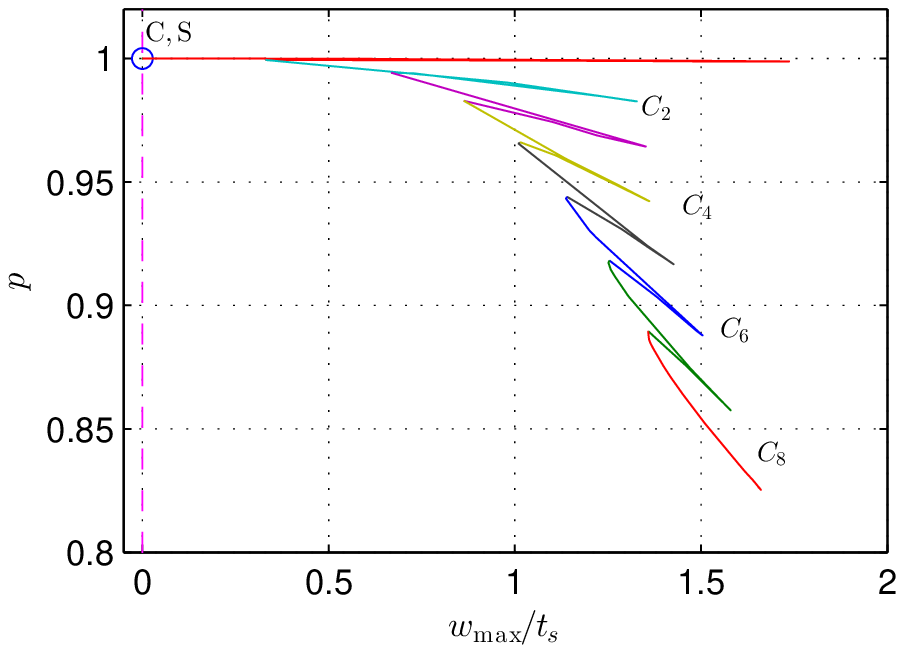}}
\subfigure[]{\includegraphics[scale=0.75]{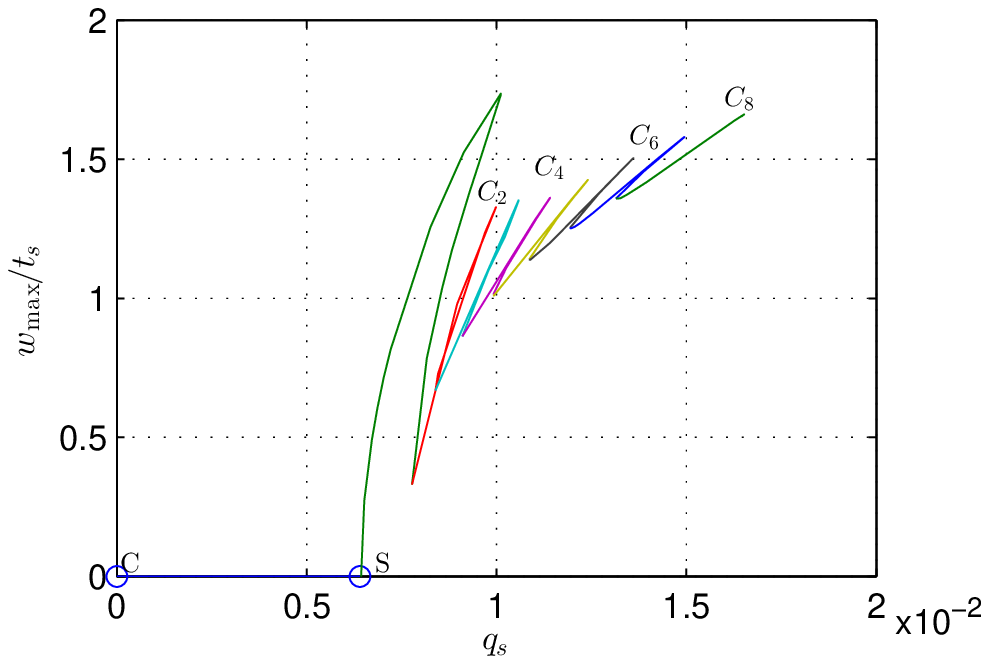}}\quad
\subfigure[]{\includegraphics[scale=0.75]{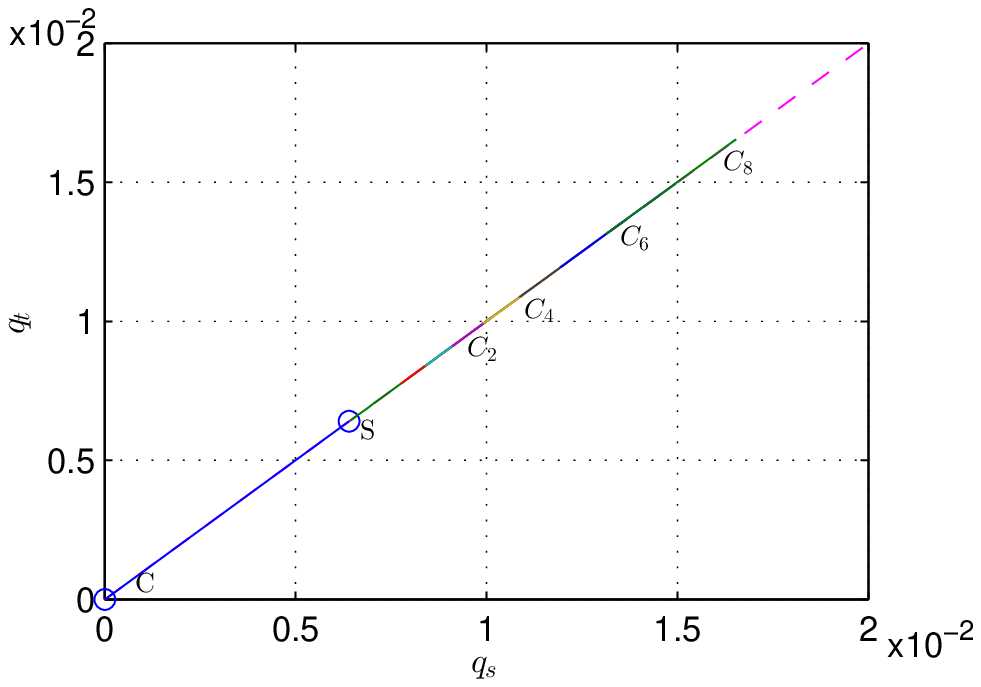}}
\caption{Results for the panel with example properties and assuming
  the main plate--stiffener joint is pinned ($c_p=0$). The normalized
  force ratio $p=P/\Pco$ is plotted versus (a) the global mode
  amplitude $q_s$ and (b) the normalized maximum out-of-plane
  deflection of the stiffener $w_\mathrm{max}/t_s$; (c) shows the
  local mode amplitude $w_\mathrm{max}/t_s$ versus the global mode
  amplitude $q_s$ in the post-buckling range; (d) shows the
  relationship between the generalized coordinates $q_t$ and $q_s$
  that define the global mode.}
  \label{fig:result}
\end{figure}

\begin{figure}[hbt]
\centering
\includegraphics[width=120mm]{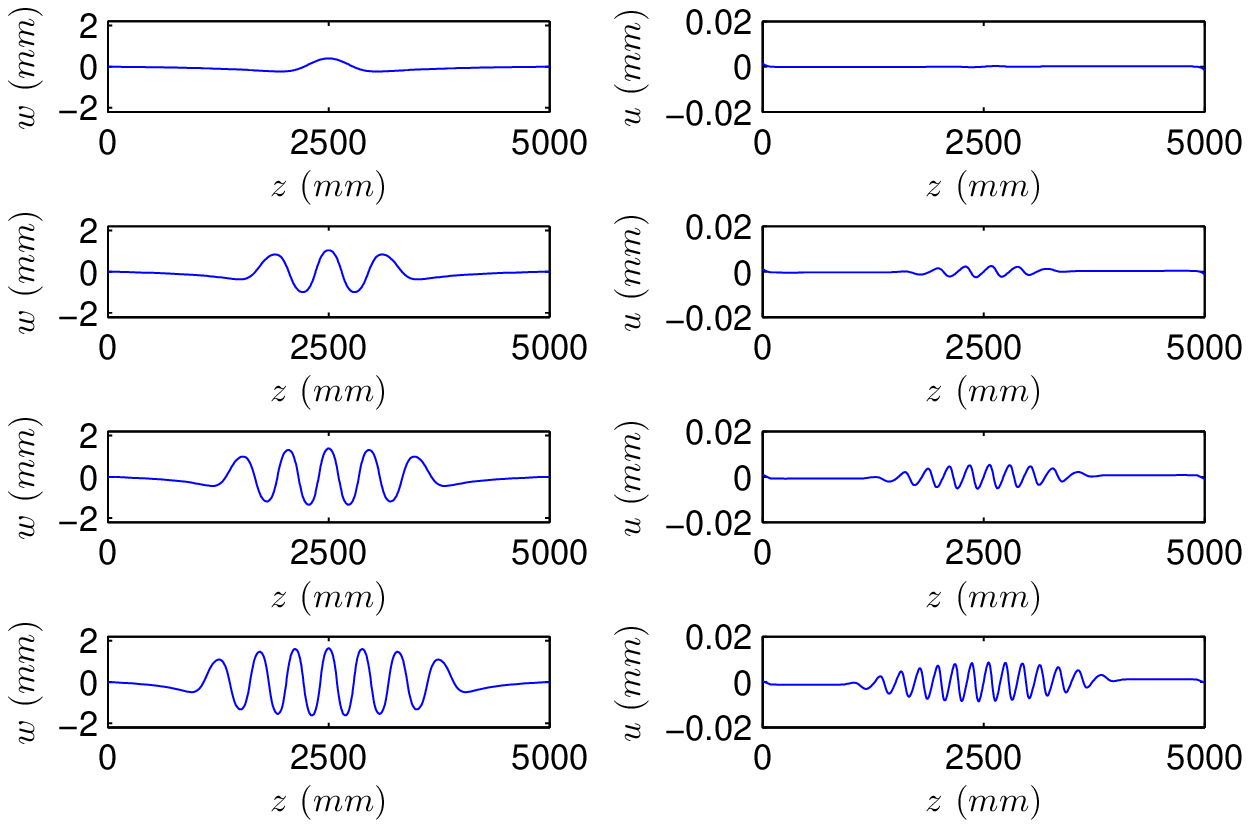}
\caption{Numerical solutions for the local out-of-plane deflection $w$
  (left) and the local in-plane deflection $u$ (right) showing the
  progressive spreading of the stiffener local buckling profile for
  cells $C_2$, $C_4$, $C_6$ and $C_8$ (top to bottom respectively)
  from the perfect model.}
\label{fig:wu_perfect}
\end{figure}

Figure \ref{fig:result}
shows the numerical results from the example properties stated at the
beginning of the current section. Initially, the results are presented
for the perfect case where the joint between the main plate and the
stiffener is pinned ($c_p=0$). The graphs in (a--b) show the
equilibrium plots of the normalized axial load $p=P/\Pco$ versus the
generalized coordinates of the sway component $q_s$ and the maximum
out-of-plane normalized deflection of the buckled stiffener
($w_\mathrm{max}/t_s$) respectively. The graph in (c) shows the
relative amplitudes of global and local buckling modes in the
post-buckling range. Finally, the graph in (d) shows the relationship
between sway $q_s$ and tilt $q_t$ components of the global buckling
mode, which are almost equal (difference approximately 0.05\%); this
indicates that the shear strain is small but, importantly, not
zero. As found in Wadee and Farsi \shortcite{WF1_2014}, for the case
where only the stiffener buckles locally, there is a sequence of
snap-backs observed. This is the signature of cellular buckling
\cite{Hunt2000} and the cells are labelled $C_i$. Figure
\ref{fig:wu_perfect}
shows the corresponding progression of the numerical solutions for the
local buckling functions $w$ and $u$ for cells $C_2$, $C_4$, $C_6$ and
$C_8$ defined in Figure \ref{fig:result}. It can be seen that the
initially localized buckling mode progressively becomes more periodic
as the system post-buckling advances.

\begin{figure}[hbt]
\centering
\subfigure[]{\includegraphics[scale=0.75]{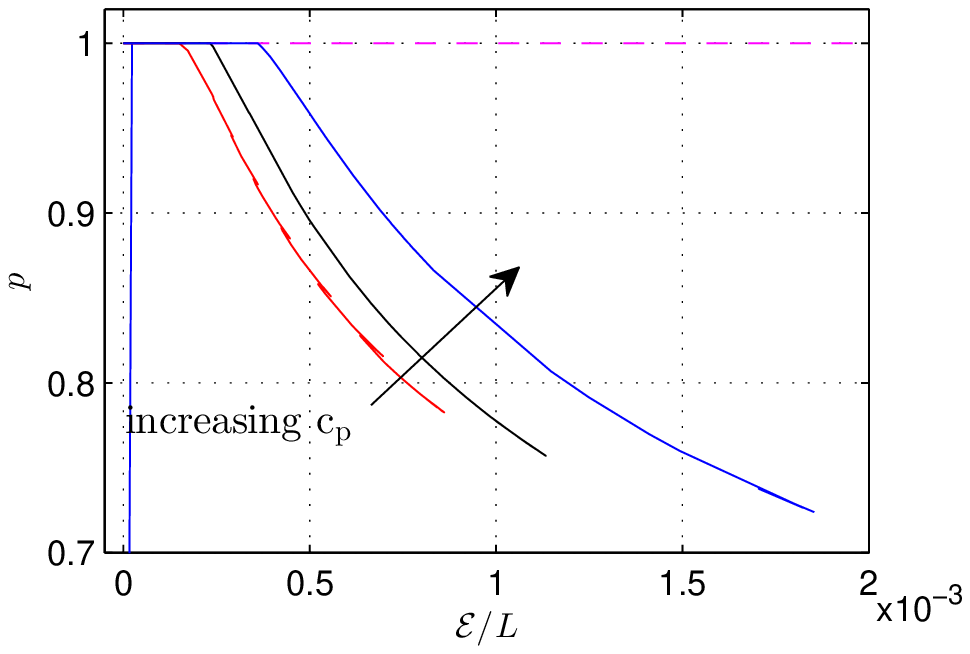}}\quad
\subfigure[]{\includegraphics[scale=0.75]{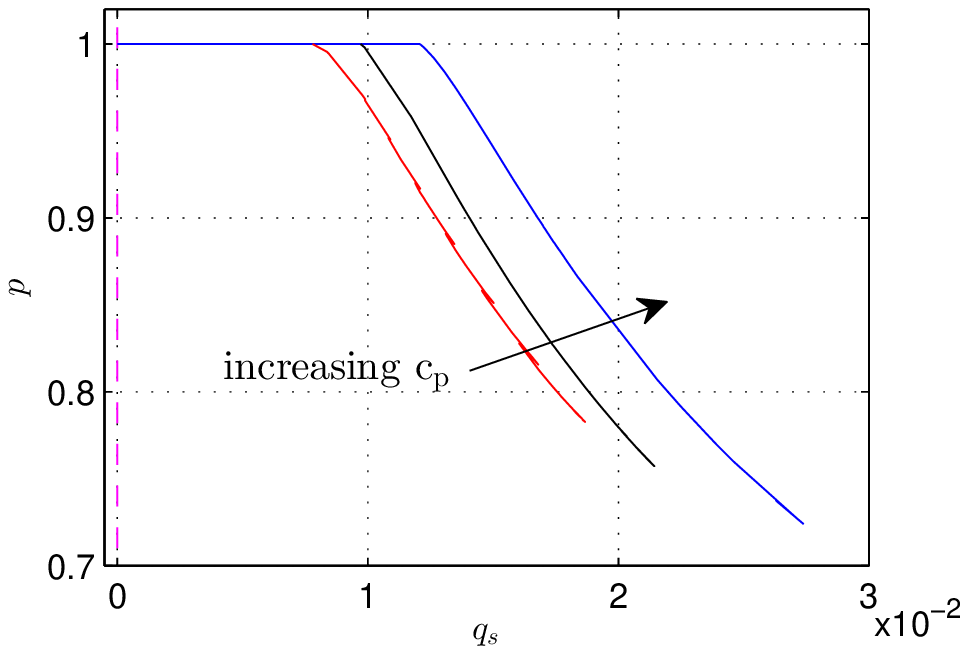}}
\subfigure[]{\includegraphics[scale=0.75]{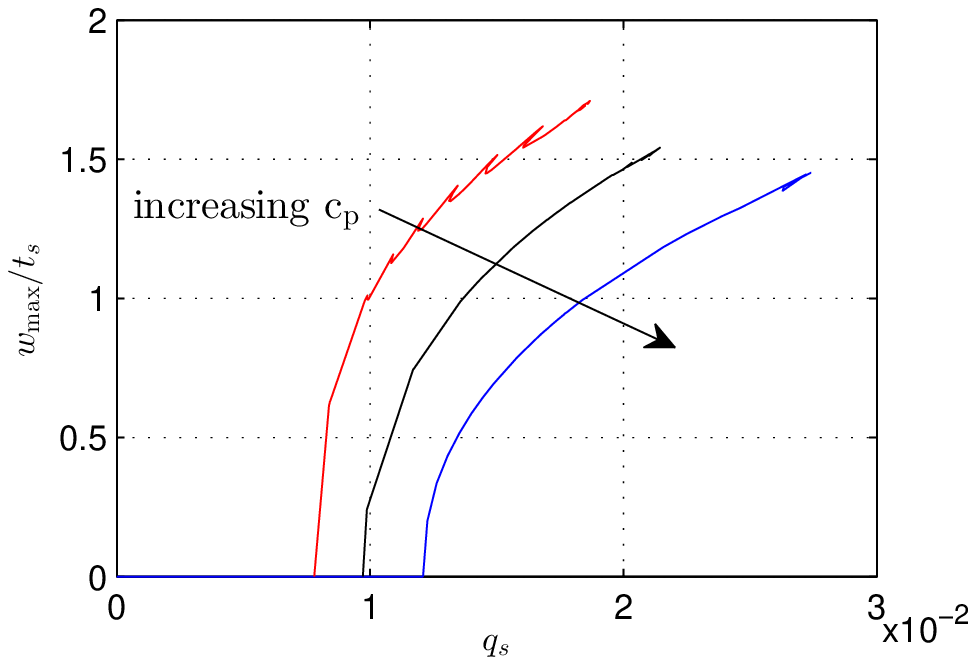}}\quad
\subfigure[]{\includegraphics[scale=0.75]{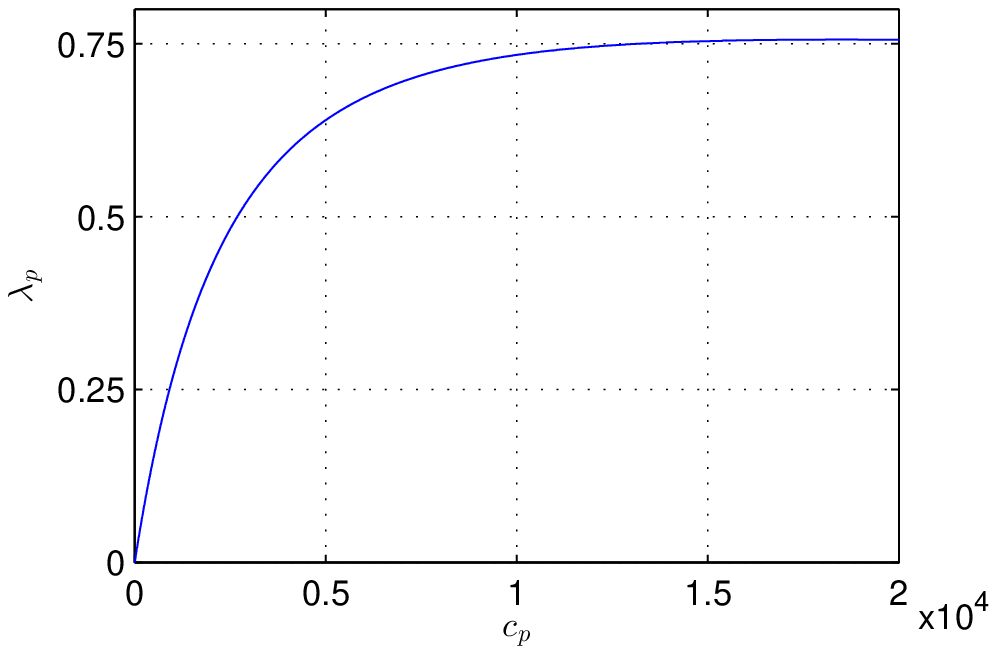}}
\caption{Variation of the equilibrium paths for increasing rigidities
  of the main plate--stiffener joint ($c_p =
  1,100,500~\mathrm{Nmm/mm}$). Graphs show the normalized force ratio
  $p$ versus (a) the normalized end-shortening, (b) the global mode
  amplitude $q_s$; (c) shows the local versus global mode
  amplitudes. (d) Plot of the ratio of the out-of-plane displacements
  in the main plate to the stiffener $\lambda_p$ versus $c_p$ in
  $\mathrm{Nmm/mm}$.}
    \label{fig:equ-qs}
\end{figure}

By comparing the equilibrium paths against those for $c_p=0$, it is
observed that the snap-backs begin to vanish as $c_p$ is
increased. Figure \ref{fig:equ-qs}
shows the equilibrium paths for the strut with an increasingly rigid
connection between the plate and the stiffener. In the graphs of the
analytical results, the values of $c_p$, in $\mathrm{Nmm/mm}$, are
taken as 1, 100 and 500 respectively. The graphs in Figure
\ref{fig:equ-qs}(a--b) show the equilibrium path of the normalized
axial load $p=P/\Pco$ versus the normalized total end-shortening
$\mathcal{E}/L$, see Equation (\ref{eq:WD}), and the global mode
amplitude $q_s$ respectively. It is found that $q_s^\mathrm{S}$
increases with $c_p$; comparing $q_s^\mathrm{S}$ for the highest $c_p$
value to the lowest value shown, it is seen to be nearly $50\%$
greater. Moreover, the post-buckling paths show a significantly
stiffer response for higher values of $c_p$. At $p = 0.81$, the
maximum out-of-plane displacement $w$ for $c_p = 0$ is approximately
$1.65t_s$, whereas for $c_p = 500~\mathrm{Nmm/mm}$, it is
approximately $1.18t_s$.  Figure \ref{fig:equ-qs}(c) shows the
normalized local versus global mode amplitudes and (d) shows the
relationship of the ratio $\lambda_p$ versus the rotational stiffness
$c_p$; the final graph shows that the relationship flattens for larger
values of $c_p$, which would be expected as a fully rigid joint
condition is approached.

\section{Validation}

The commercial FE software package \textsc{Abaqus} \shortcite{ABAQUS} was
first employed to validate the results from the analytical model. The
same example set of section and material properties were chosen, as
presented in \S\ref{sec:numerical}.  Four-noded shell elements with
reduced integration (S4R) were used to model the structure. Rotational
springs were also used along the length to simulate the rigidity of
the main plate--stiffener joint. An eigenvalue analysis was used to
calculate the critical buckling loads and eigenmodes. The nonlinear
post-buckling analysis was performed with the static Riks method
\cite{Riks1972} with the aforementioned eigenmodes being used to
introduce the necessary geometric imperfection to facilitate this. In
the current example, the rotational spring stiffness $c_p$ is assumed
to be $1000~\mathrm{Nmm/mm}$, which gives a value of $\lambda_p =
0.2687$ and gives negligible rotation at the main plate--stiffener
connection.  Linear buckling analysis shows that global buckling is
the first eigenmode; Table \ref{tab:plpc}
\begin{table}[htb]
\centering
\begin{tabular}{ccccc}
  Source & $\sigma_o^\mathrm{C}~(\mathrm{N/mm^2})$ &
  $\sigma_{l,s}^\mathrm{C}~(\mathrm{N/mm^2})$ &
  $\sigma_{l,p}^\mathrm{C}~(\mathrm{N/mm^2})$ &
  Critical mode\\
  \hline
  Theory & $4.948 $ & $236.02$ & $539.91$ & Global\\
  FE & $4.942$ & $228.94$ & $503.15$ & Global\\
  \hline
  \% difference & 0.12 & 3.09 & 7.31 & N/A \\
  \hline
\end{tabular}
\caption{Theoretical and FE values of the global and local critical buckling
  stresses ($\sigma_o^\mathrm{C}$ and $\sigma_l^\mathrm{C}$)
  respectively; subscripts ``$p$'' and ``$s$'' refer to the main plate
  and the stiffener respectively and $k_p$ is taken to be 1.247 for
  the stiffener and 6.97 for the main plate (assuming a rigid joint
  exists at the main plate--stiffener connection). The expression for
  $\sigma_o^\mathrm{C}=\Pco/A$, where $A$ is the cross-sectional area
  of the strut.}
\label{tab:plpc}
\end{table}
presents the critical stresses for all the components from the
analytical and the FE models.

\begin{figure}[hbtp]
\centering
\subfigure[]{\includegraphics[scale=0.8]{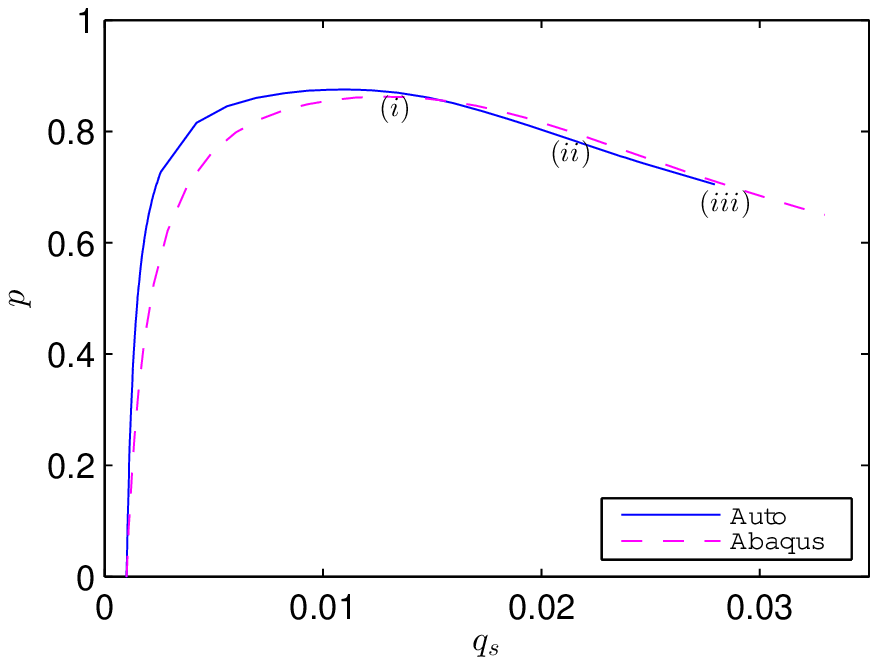}}\quad
\subfigure[]{\includegraphics[scale=0.8]{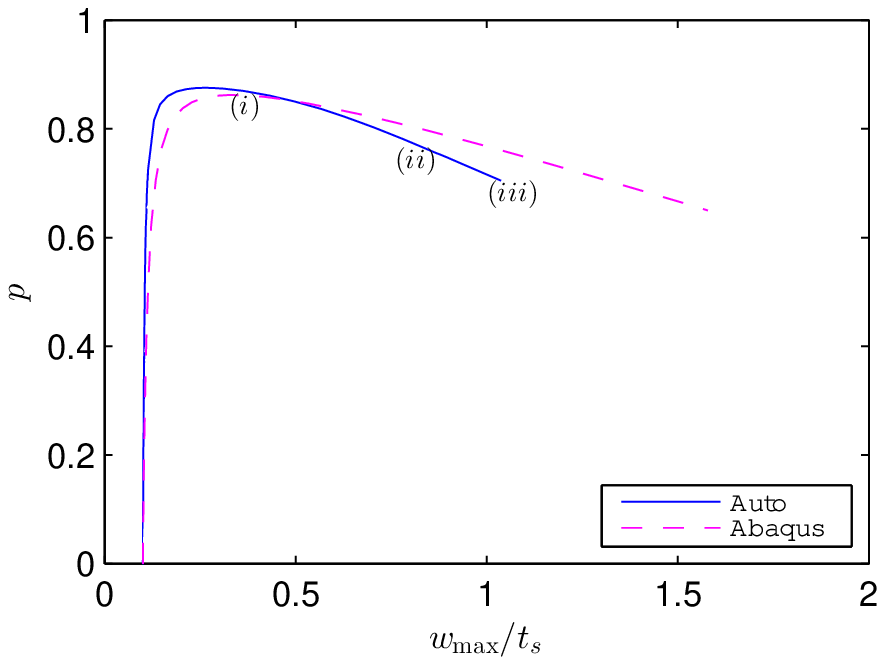}}
\subfigure[]{\includegraphics[scale=0.8]{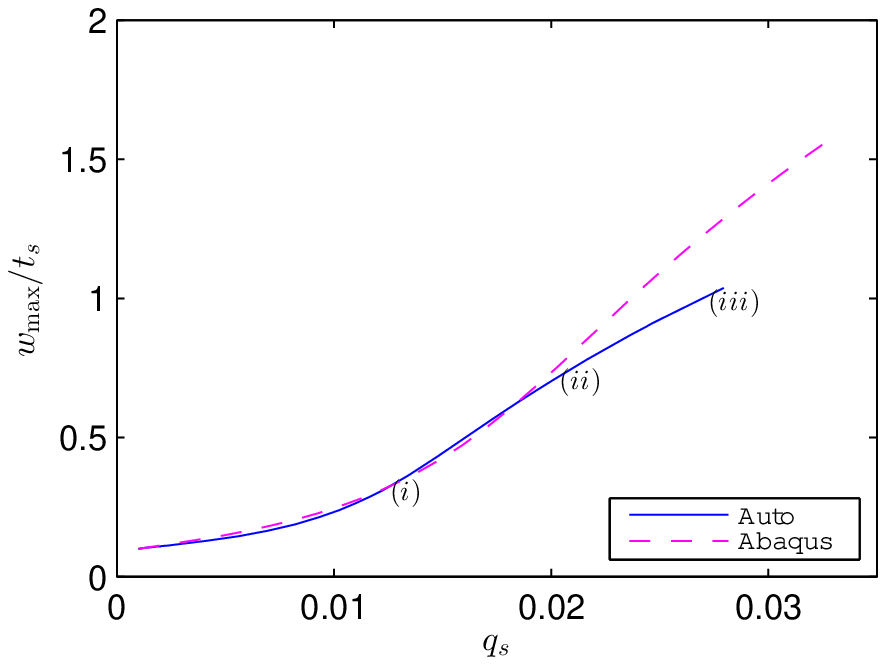}}
\caption{Comparisons of the analytical model (solid lines) versus the
  FE model (dashed lines) solutions, both with $c_p =
  1000~\mathrm{Nmm/mm}$. The plots show the normalized force ratio $p$
  versus (a) the global mode amplitude $q_s$ and (b) the maximum
  out-of-plane normalized stiffener deflection $w_\mathrm{max}/t_s$;
  (c) local versus global mode amplitudes.}
  \label{fig:valid_abaqus}
\end{figure}

\begin{figure}[hbtp]
\centering
\includegraphics[scale=1.0]{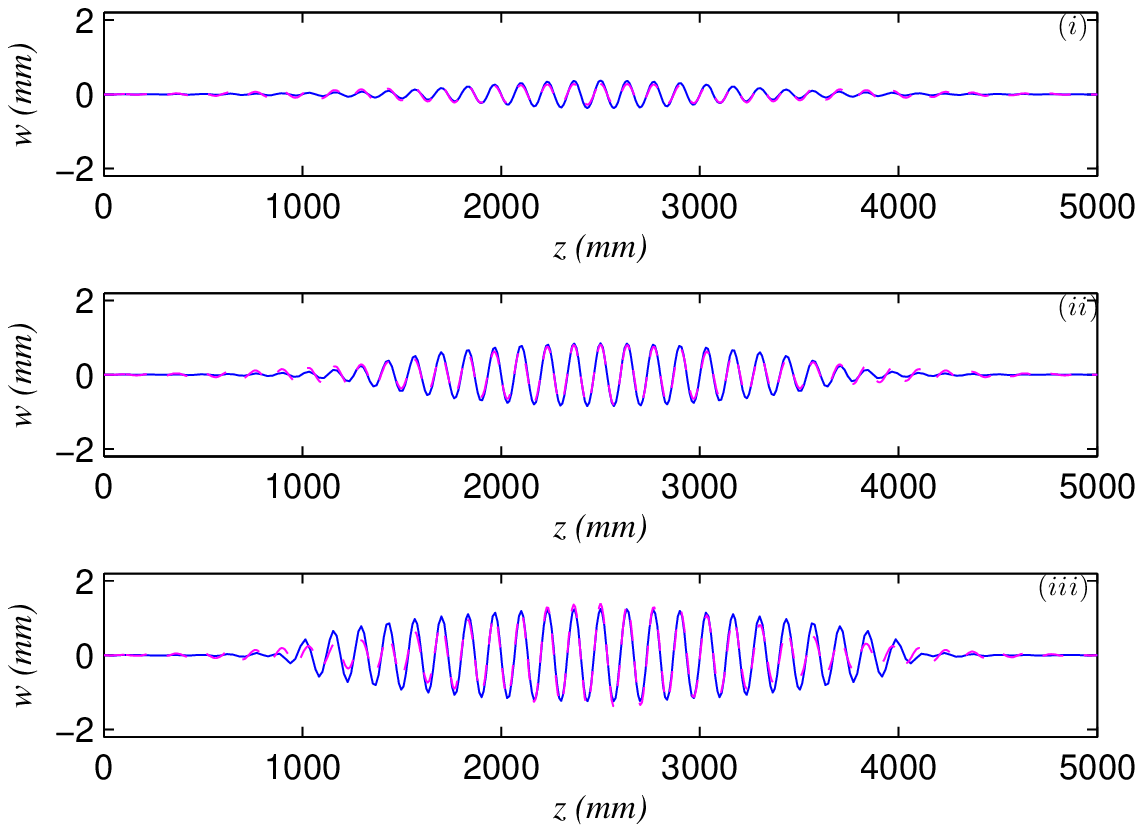}
\caption{Comparisons of the analytical model (solid lines) versus the
  FE model (dashed lines) solutions for the out-of-plane deflection of
  the stiffener $w$ for the points shown in (i)--(iii).}
  \label{fig:valid-arcl}
\end{figure}

\begin{figure}[hbtp]
\centering
\subfigure[]{\includegraphics[scale=0.65]{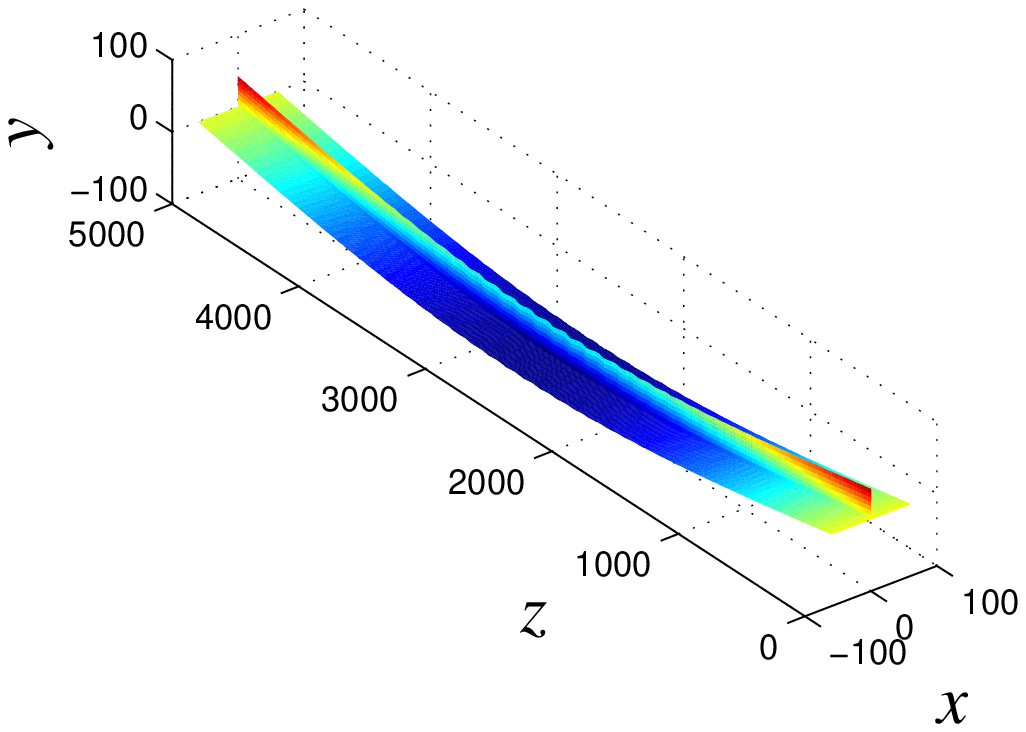}\quad
  \includegraphics[scale=0.65]{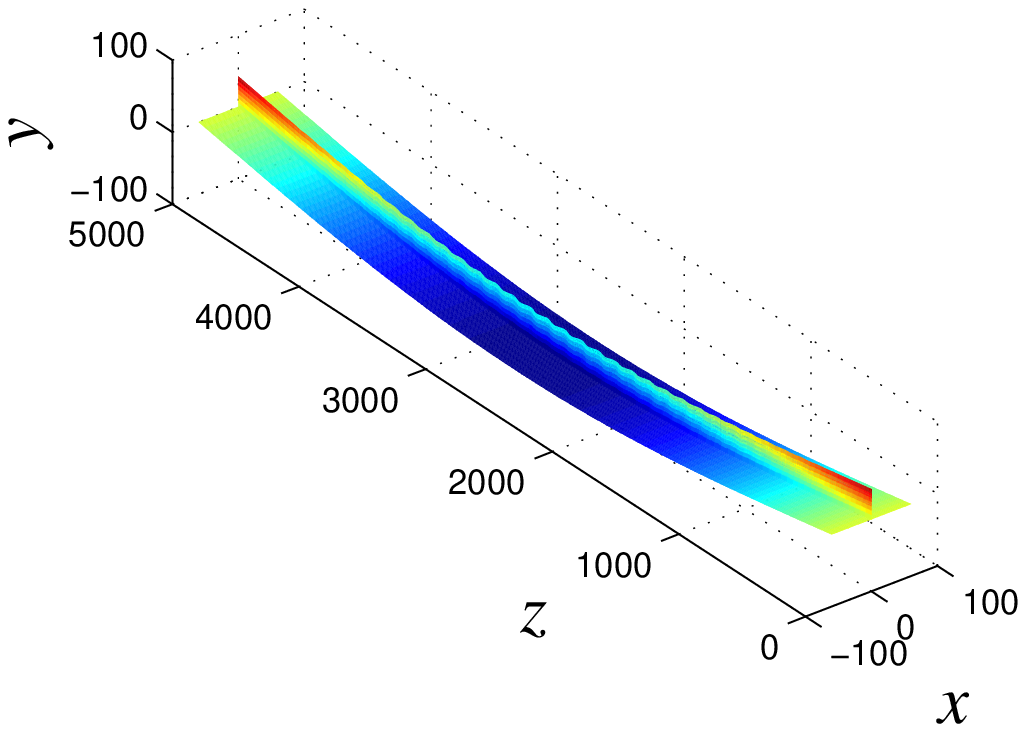}}
\subfigure[]{\includegraphics[scale=0.65]{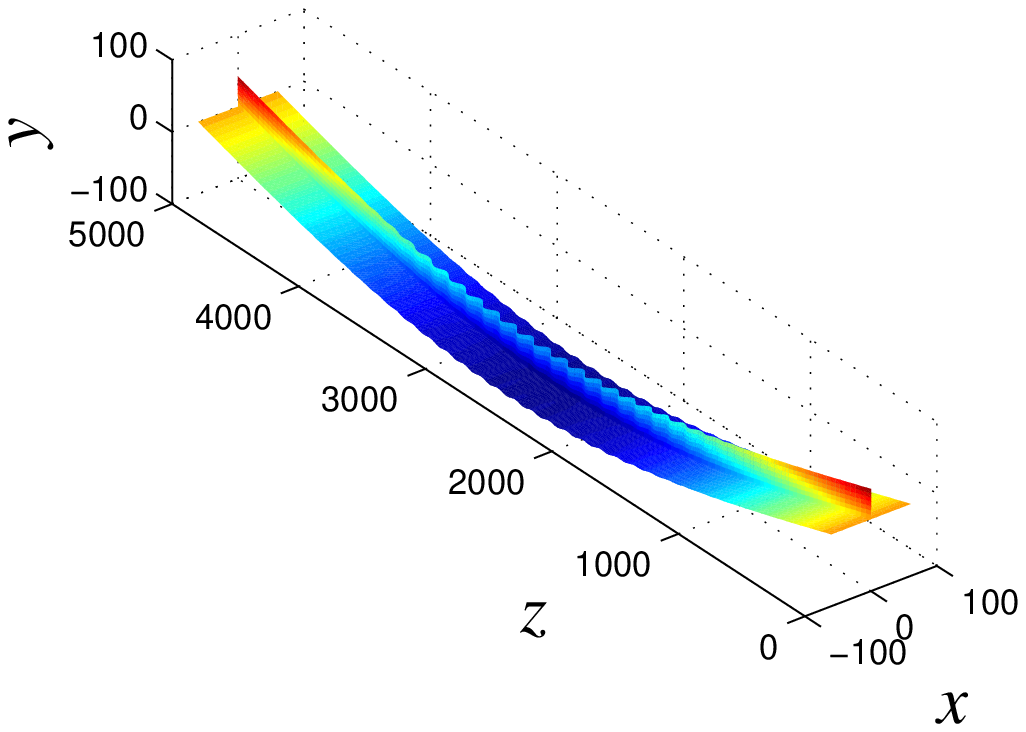}\quad
  \includegraphics[scale=0.65]{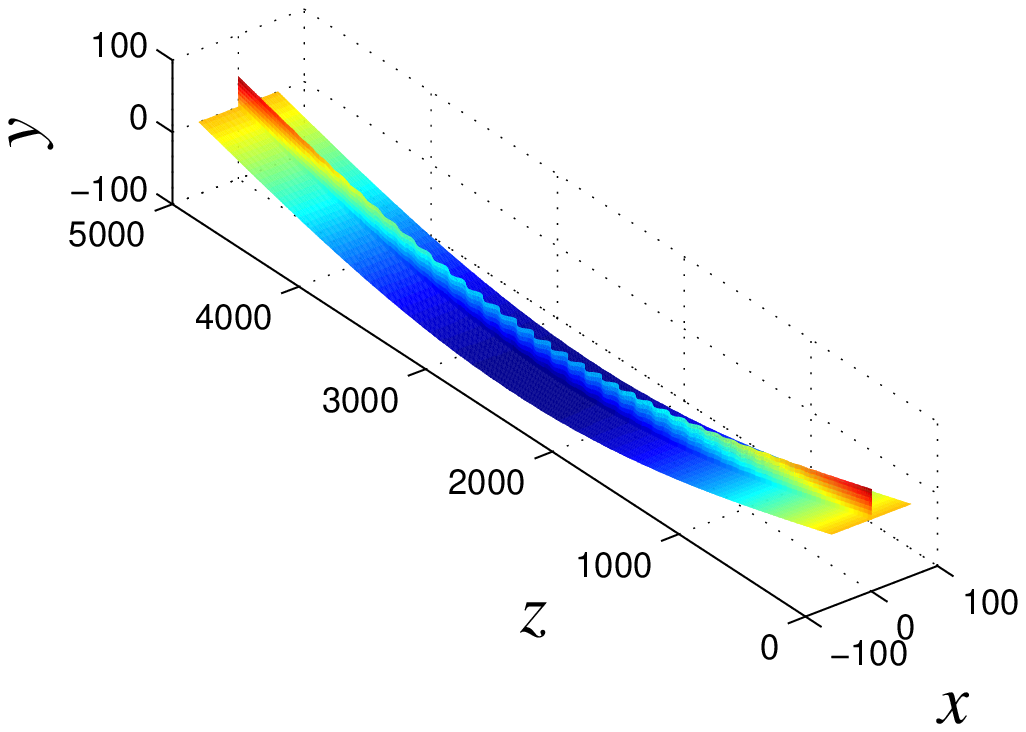}}
\subfigure[]{\includegraphics[scale=0.65]{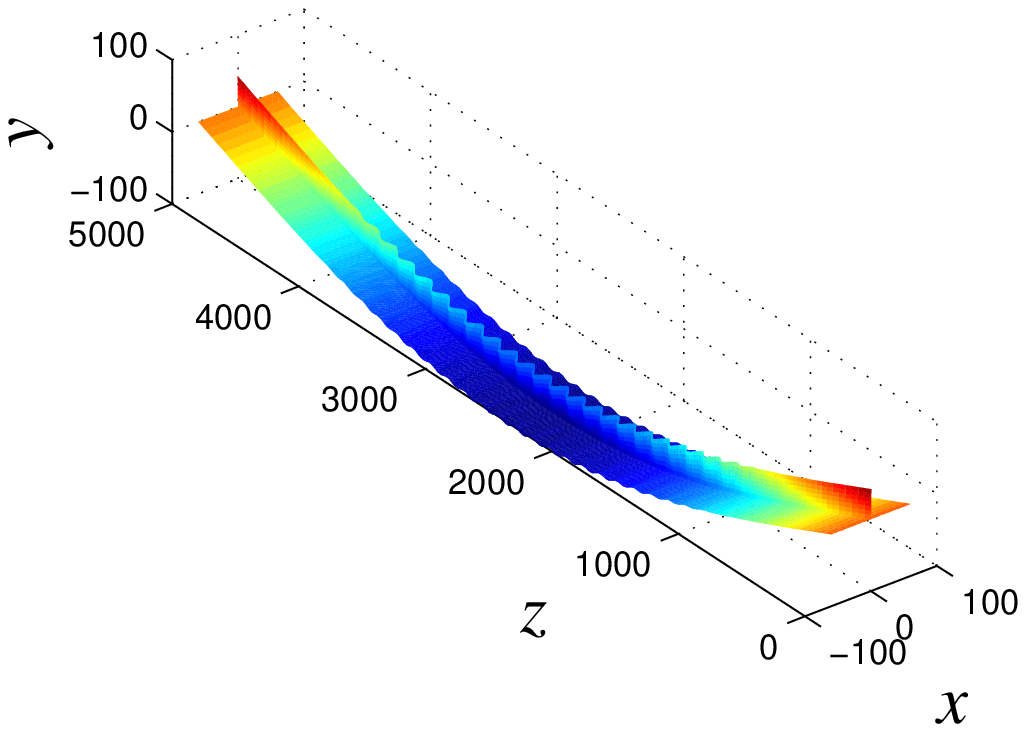}\quad
  \includegraphics[scale=0.65]{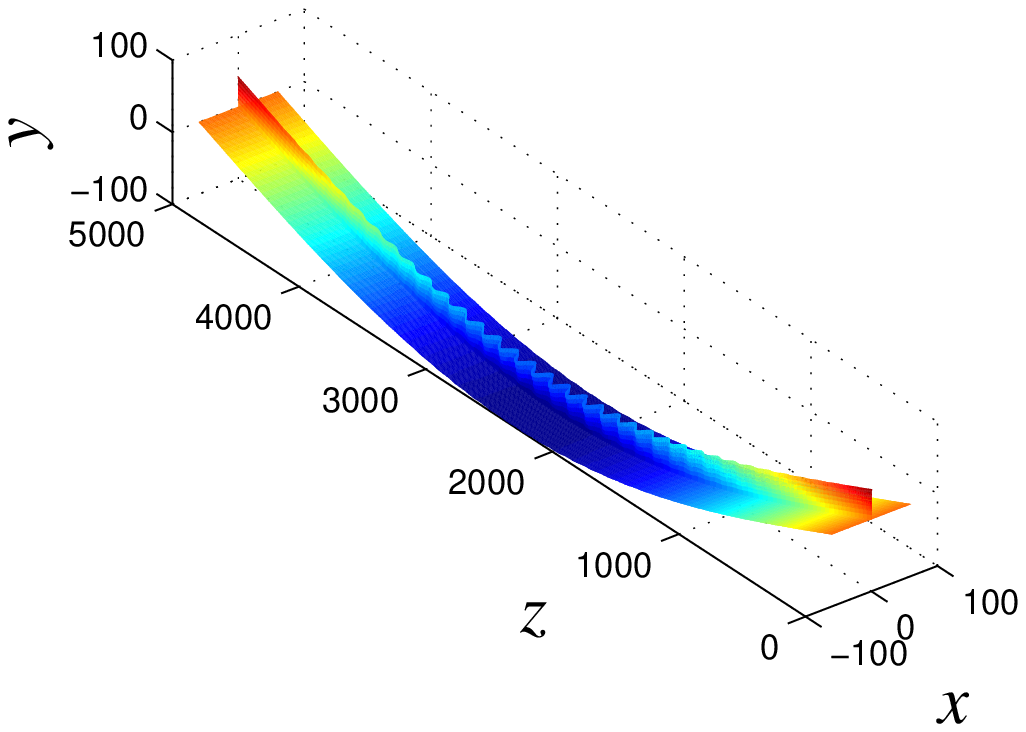}}
\caption{Comparisons of the numerical solutions from (left) the
  analytical model and (right) the FE model with
  $c_p=1000~\mathrm{Nmm/mm}$ visualized on 3-dimensional
  representations of the strut. The results are shown for equilibrium
  states at the points shown in Figure
  \protect\ref{fig:valid_abaqus}(a--c) defined as (i)--(iii). All
  dimensions are in millimetres but the local buckling displacements
  are scaled by a factor of 2 to aid visualization.}
\label{fig:valid-3d}
\end{figure}

Figure \ref{fig:valid_abaqus}
shows the comparisons between the numerical results from the
analytical and the FE models. It is shown for the
case with an initial global imperfection $q_{s0}=0.001$ and a local
imperfection, where $A_0=0.12~\mathrm{mm}$, $\alpha=5.0$, $\beta=75$
and $\eta=L/2$ for $w_0$, which represents the initial imperfection
for the analytical model that matches the FE model imperfection
satisfactorily such that a meaningful comparison can be made. The
graphs in (a--b) show the normalized axial load $p$ versus the global
and the local mode amplitudes respectively; the graph in (c) shows the
local versus the global buckling modal amplitudes. Figure
\ref{fig:valid-arcl}
shows the local out-of-plane deflection profiles $w$ at the respective
locations (i)--(iii), shown in Figure \ref{fig:valid_abaqus}(a--c),
the comparison being for the same value of $p$. As can be seen, both
sets of graphs show excellent correlation in all aspects of the
mechanical response; in particular, the results for $w$ from the
analytical and the FE model are almost indistinguishable. This shows a
marked improvement on the previous simpler model \cite{WF1_2014} where
the main plate was assumed not to buckle locally. A visual comparison
between the 3-dimensional representations of the strut from the
analytical and the FE models is also presented in Figure
\ref{fig:valid-3d}.

\subsection{Comparison with experimental study}

An experimental study of a thin-walled stiffened plate conducted by
Fok \emph{et al.}\ \shortcite{Fok1976} also focused on the case where
global buckling was critical. Two specific tests were conducted on a
panel with multiple stiffeners. The experimental results were compared
with the current analytical model and also with the FE model
formulated in \textsc{Abaqus}.  The cross-section of the stiffened
panel investigated is shown in Figure \ref{fig:panel}.  The
experimental panel had $10$ blade stiffeners with spacing
$b=45.5~\mm$, height $h_1=13.5~\mm$, thicknesses $t_s=t_p=0.735~\mm$
and $h_2=t_p/2$ (\emph{i.e.}\ stiffeners on one side only). The
experimental specimen was constructed from cold-setting
Araldite\textsuperscript{\textregistered} (epoxy resin) and the
material had an elastic stress--strain relationship, but no material
properties were provided. Hence, in the analytical and FE models,
nominal values of $E$ and $\nu$ were used ($E=210~\mathrm{kN/mm^2}$
and $\nu=0.3$ as before); this did not pose a problem so long as the
same values were used in both models. Moreover, it is worth noting
that the behaviour of the experimental specimens was reported to be
elastic and only ratios of loads and displacements were reported as
the results \cite{Fok1976}.

In the first test, the length $L$ of the panel was $400~\mm$ ensuring
that the global critical buckling load was much less than the local
buckling load. The initial global imperfection was measured to be
$W_0=1.2t_s$ but there were negligible out-of-plane imperfections in
the stiffeners and in the main plate (\emph{i.e.}\ $w_0=w_{p0}=0$ were
assumed).  In the second test, the length of the stiffened panel was
reduced to $L=320~\mm$ with the consequence that the local buckling
stress was only approximately 5\% above that for the global mode. The
corresponding critical stresses for both tests are summarized in Table
\ref{tab:fok-pc}.
\begin{table}[tbhp]
\centering
\begin{tabular}{ccccc}
  & $L~(\mm)$ & $\sigma_o^\mathrm{C}~(\mathrm{N/mm^2})$ &
  $\sigma_{l,s}^\mathrm{C}~(\mathrm{N/mm^2})$ &
  $\sigma_{l,p}^\mathrm{C}~(\mathrm{N/mm^2})$\\
  \hline
  Test 1 & $400$ & $7.122 \times 10^{-4} E$ & $3.34 \times 10^{-3} E$
  & $1.176 \times 10^{-3} E$ \\
  Test 2 & $320$ & $1.109 \times 10^{-3} E$ & $3.34 \times 10^{-3} E$
  & $1.176 \times 10^{-3} E$ \\
  \hline
\end{tabular}
\caption{Theoretical values of the global and local critical
  buckling stresses ($\sigma_o^\mathrm{C}$ and $\sigma_l^\mathrm{C}$)
  respectively; subscripts ``$p$'' and ``$s$'' refer to the main plate
  and the stiffener respectively. The expression for
  $\sigma_o^\mathrm{C}=\Pco/A$, where $A$ is the cross-sectional area
  of the panel.}
\label{tab:fok-pc}
\end{table}

\begin{figure}[hbtp]
\centering
\subfigure[]{\includegraphics[scale=0.7]{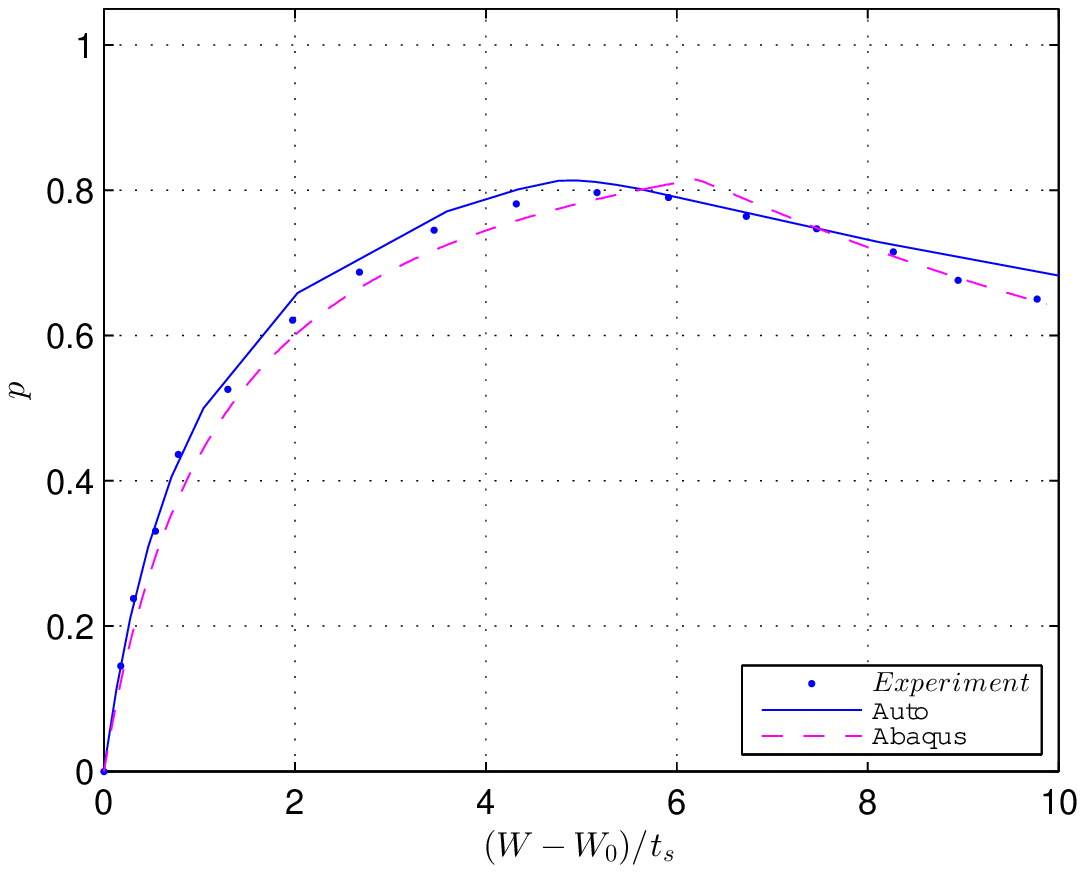}}\quad
\subfigure[]{\includegraphics[scale=0.7]{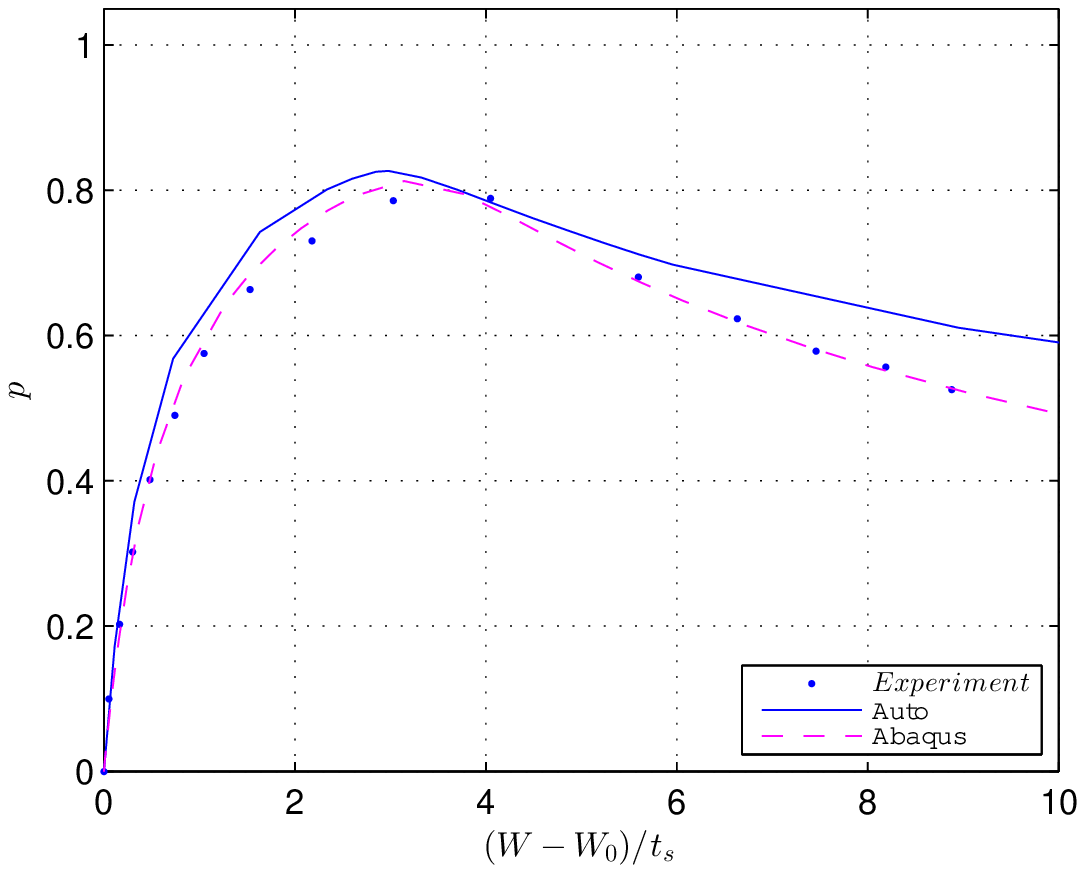}}
\caption{Comparisons of the experimental results (dots) with the
  solutions from the analytical model (solid lines) and the FE model
  (dashed lines). Graphs of the normalized force ratio $p$ versus the
  normalized global mode deflection relative to the initially
  imperfect state for the panel $(W-W_0)/t_s$ are shown for (a) Test 1
  where $L=400~\mm$ and (b) Test 2 where $L=320~\mm$.}
  \label{fig:valid_exp}
\end{figure}

\begin{figure}[hbtp]
\centering
\subfigure[]{\includegraphics[scale=0.7]{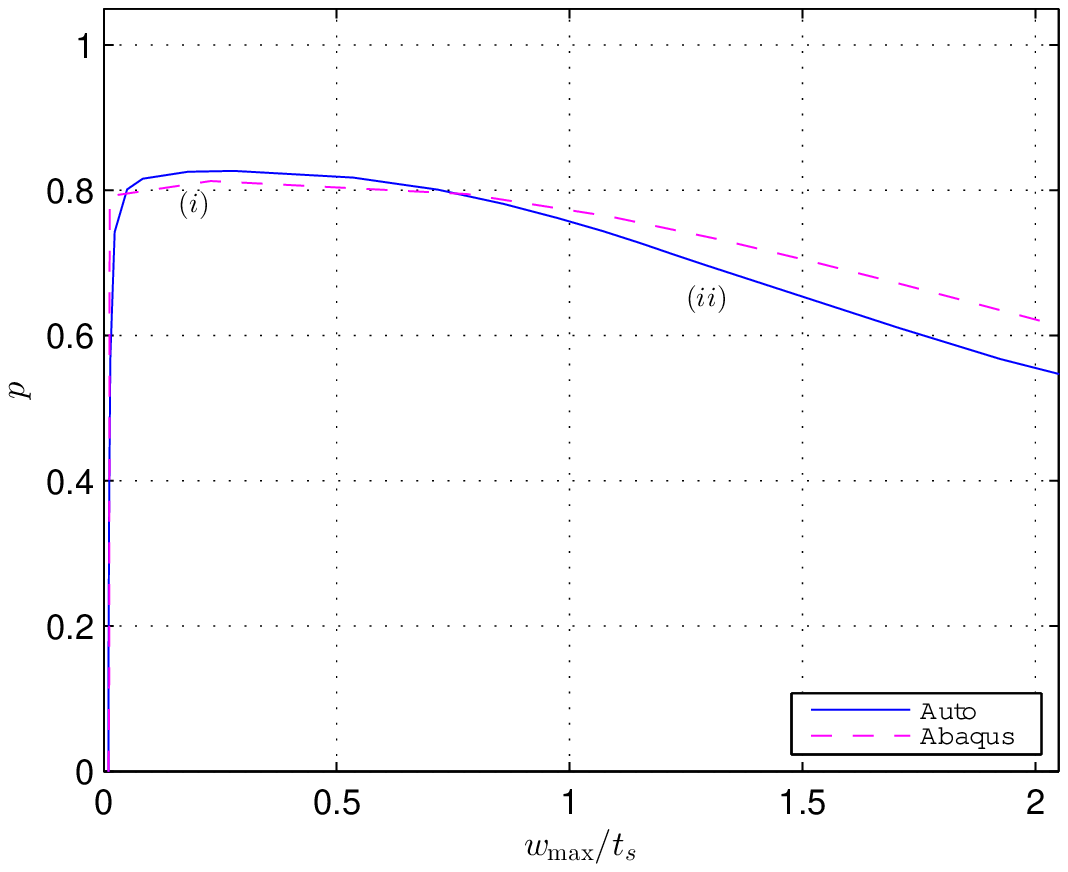}}\quad
\subfigure[]{\includegraphics[scale=0.7]{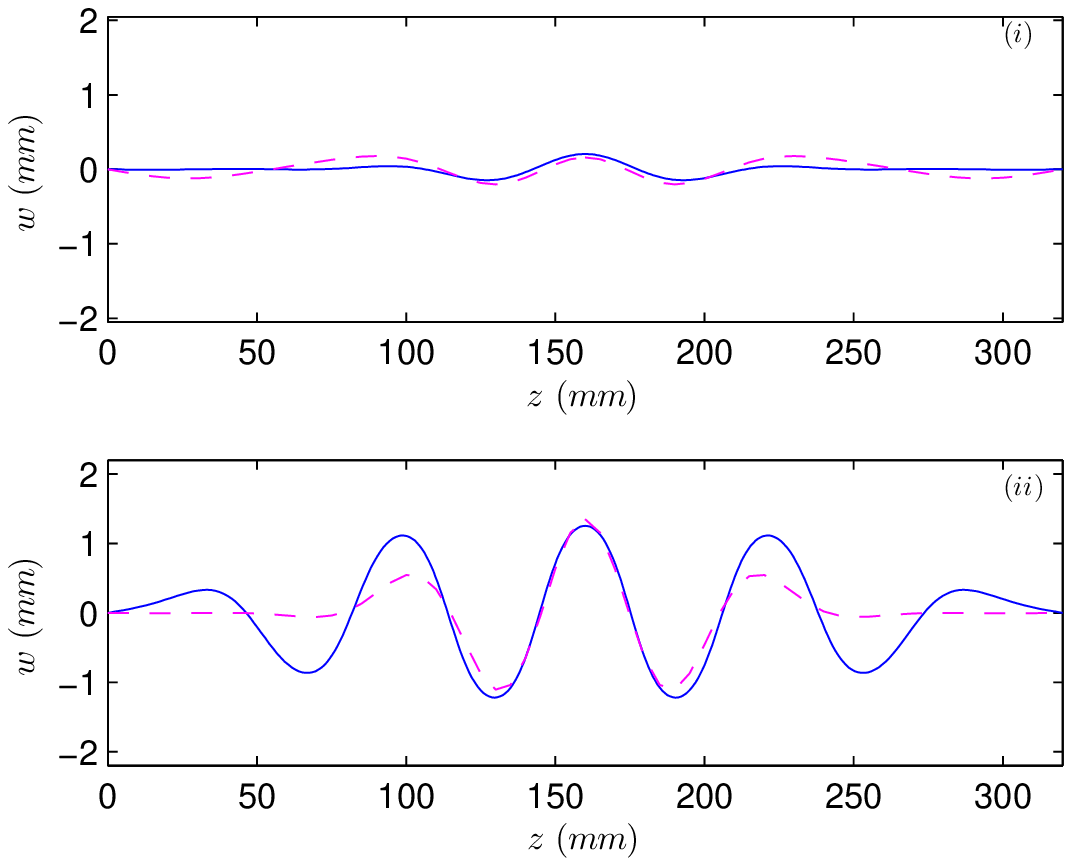}}
\caption{Comparison of the analytical results (solid lines) with the
  FE model (dashed lines) for the Test 2 properties. (a) Plot of the
  normalized force ratio $p=P/\Pco$ versus the normalized maximum
  out-of-plane deflection $w_\mathrm{max}/t_s$. (b) Local buckling
  out-of-plane deflections $w$ of the stiffener for the points shown
  in (i) and (ii).}
  \label{fig:fok_arcl}
\end{figure}

For the $L=320~\mm$ panel, the initial global buckling mode
imperfection $W_0$ was set to $0.8t_s$ and the amplitude of the
out-of-plane imperfection $A_0=0.01t_s$ with $\alpha=4$, $\beta=11$
and $\eta=L/2$. For both models, analytical and FE, the stiffness of
the rotational spring $c_p$ was calibrated to be $300~\mathrm{Nmm/mm}$
since this gave the best match with the peak load of the experimental
results.  To find the equilibrium path, the numerical continuation
process was initiated from zero load. Figure \ref{fig:valid_exp}
shows the comparisons between the experimental results from Fok
\emph{et al.}\ \shortcite{Fok1976}, the analytical and the FE
models. The comparisons show strong agreement between all three sets
of results.  Since there was no information provided about the local
out-of-plane deflection magnitude, the results from the analytical
model are compared to the FE results directly. Figure
\ref{fig:fok_arcl}(a)
presents the comparison of the normalized force ratio $p=P/\Pco$
versus the maximum normalized out-of-plane stiffener deflection
$w_\mathrm{max}/t_s$, where the initial global and local imperfection
sizes and shapes were given as before.  Figure \ref{fig:fok_arcl}(b)
shows the comparisons of the analytical with the FE model results for
the local out-of-plane displacement of the stiffener $w$ for the Test
2 properties. Note that the results are obtained when (i) $p=0.80$ and
(ii) $p=0.65$. The comparison between the local buckling amplitudes
and wavelength is excellent. Of course at lower loads, in the advanced
post-buckling range, there is divergence between the non-midspan
peaks; this is a further example of the FE model locking the modal
wavelength as found in earlier studies even though actual experimental
evidence shows the contrary \cite{Becque_thesis,WG2012,WB2014,WF1_2014}.

\section{Concluding remarks}

An analytical model based on variational principles has been extended
to model local--global mode interaction in a stiffened plate subjected
to uniaxial compression. By introducing the sympathetic deflection of
the main plate along with the locally buckling stiffener, the current
model could now be compared to published experiments \cite{Fok1976}
and a finite element model formulated in \textsc{Abaqus}; results from
both are found to be excellent. Currently, the authors are conducting
an imperfection sensitivity study to quantify the parametric space for
designers to avoid such dangerous structural behaviour, the results of
which would provide greater understanding of the interactive buckling
phenomena and highlight the practical implications.

\bibliography{refs}

\end{document}